\documentclass[reqno,12pt]{amsart}
\usepackage{fullpage}
\usepackage{amsfonts}
\usepackage{amssymb}
\usepackage{latexsym}
\usepackage{mathrsfs}
\usepackage{multicol}
\usepackage{delarray}
\numberwithin{equation}{section}

\newtheorem{theorem}{Theorem}[section]{\bf}{\it}

{\bf}{\it}

{\bf}{\it}

\newtheorem{proposition}[theorem]{Proposition}{\bf}{\it}

\newcommand{\p}{\partial}
\newcommand{\mtb}{\mathbf}
\newcommand{\f}{\frac}
\newcommand{\tf}{\tfrac}

\def\tr{{\rm tr}}
\def\eos/{equation of state}
\def\esos/{equations of state}
\def\com/{constant of motion}
\def\csom/{constants of motion}

\begin{document}

\title{Conservation laws of inviscid non-isentropic compressible fluid flow in $n>1$ spatial dimensions}

\author{
Stephen C. Anco${}^1$ 
\lowercase{\scshape{and}} 
Amanullah Dar${}^{1,2}$ \\
\\\lowercase{\scshape{
${}^1$Department of Mathematics, Brock University, 
St. Catharines, ON Canada}}\\
\\\lowercase{\scshape{
${}^2$Department of Mathematics, Quaid-e-Azam University, 
Islamabad, Pakistan }}
}

%\address{Department of Mathematics, Brock University, St. Catharines, ON L2S3A1 Canada }
%\curraddr{Department of Mathematics, Quaid-e-Azam University, Islamabad, Pakistan }

\email{sanco@brocku.ca}
\email{amanullahdar@hotmail.com}

\thanks{S.C.A. is supported by an NSERC research grant. A.D. thanks HEC, Pakistan, for providing a 6-month fellowship grant and the Department of Mathematics at Brock University for additional support during the extended period of a visit when this research was completed.}

\begin{abstract}
Recent work giving a classification of kinematic and vorticity 
conservation laws of compressible fluid flow with barotropic equations of state
(where pressure is a function only of the fluid density) 
in $n>1$ spatial dimensions 
is extended to general non-isentropic equations of state 
in which the pressure is also a function of 
the dynamical entropy (per unit mass) of the fluid. 
Two main results are obtained. 
First, 
we find that apart from the familiar conserved integrals 
for mass, momentum, energy, angular momentum and Galilean momentum, 
and volumetric entropy, 
additional kinematic conserved integrals 
arise only for non-isentropic equations of state given by 
a generalized form of the well-known polytropic equation of state 
with dimension-dependent exponent $\gamma=1+2/n$,
such that the proportionality coefficient is an arbitrary function of 
the entropy (per unit mass). 
Second, we show that 
the only vorticity conserved integrals consist of a circulatory entropy
(which vanishes precisely when the fluid flow is irrotational)
in all even dimensions.
In particular, 
the vorticity integrals for helicity in odd dimensions 
and enstrophy in even dimensions 
are found to be no longer conserved for any non-isentropic equation of state. 
\end{abstract}
\maketitle

\section{Introduction and summary}

The mathematical study of $n$-dimensional fluid flow
has attracted rising interest in the past few decades, 
encompassing work on symmetries and conservation laws \cite{Ovs,Ibr1973,Ibr},
Hamiltonian structures \cite{Ver}, Casimir invariants \cite{Ser,Dez,KheChe},
and other group-theoretic aspects of the $n$-dimensional
Eulerian fluid equations \cite{Arn1966,Arn1969,ArnKhe}.
Further results in the special case of $n=1$ dimension appear in 
\cite{Ver1,Ver2,OlvNut}.

In a recent contribution \cite{paperI}, 
we undertook a systematic study of local conservation laws 
for the Euler equations governing isentropic compressible fluid flow 
in $n>1$ spatial dimensions,
where the pressure of the fluid is a function of the fluid density
as given by some barotropic \eos/, 
while the entropy (per unit mass) of the fluid is constant 
throughout the fluid domain. 
Our results fully settled the problem of finding 
all $n$-dimensional local continuity equations 
in two cases of primary interest: 
{\em kinematic} conservation laws, 
like mass, momentum and energy, 
for which the conserved density and spatial flux involve only
the fluid velocity, density and pressure, 
in addition to the time and space coordinates; 
and {\em vorticity} conservation laws,
such as helicity in three dimensions 
as well as circulation and enstrophy in two dimensions, 
where the conserved density and spatial flux have an essential dependence on 
the curl of the fluid velocity. 
These two classes of conservation laws comprise 
all of the local continuity equations known to-date 
for isentropic compressible fluid flow in $n>1$ dimensions. 

The present paper extends this work to 
non-isentropic adiabatic compressible fluid flow
in which the entropy (per unit mass) is conserved only along streamlines 
and the pressure is given by an \eos/ in terms of 
both the fluid density and entropy. 
In particular, 
we explicitly derive all $n$-dimensional kinematic and vorticity
conservation laws, including any that are admitted only for 
special non-isentropic \esos/ or in special dimensions $n>1$,
with the conserved densities and spatial fluxes allowed to depend on 
the entropy (in addition to the previous fluid variables). 

In section~\ref{kin_conslaws}, 
we begin by reviewing the formulation of necessary and sufficient equations 
for directly determining the conserved densities admitted by 
the Euler equations for $n$-dimensional inviscid non-isentropic compressible fluid flow.
We next verify that the physically familiar kinematic conserved integrals
for  mass, momentum, energy, angular momentum and Galilean momentum 
in compressible fluid flow with a general isentropic \eos/ 
remain conserved for the $n$-dimensional non-isentropic Euler equations. 
By solving the determining equations for kinematic conserved densities
in $n>1$ dimensions, 
we then show that the only additional conserved integrals 
consist of volumetric entropy in a generalized form, 
plus two generalized energies arising for polytropic \esos/
where the pressure is proportional to 
a particular dimension-dependent power $\gamma =1+2/n$ of the density, 
with the proportionality coefficient given by an arbitrary function of 
the entropy (per unit mass). 

In section~\ref{vort_conslaws}, 
we first consider the odd-dimensional helicity integral
and the even-dimensional enstrophy integral 
both of which are known to be conserved in 
$n$-dimensional isentropic fluid flow \cite{KheChe,paperI}. 
By examining how the corresponding local vorticity conservation laws 
depend on the \eos/ of the fluid, 
we show that the helicity and enstrophy integrals are no longer conserved
for any non-isentropic \esos/. 
Next we solve the determining equations to see 
if there are any vorticity conserved densities in $n>1$ dimensions 
admitted by the non-isentropic Euler equations. 
As a main new result, 
this classification yields a non-trivial conserved integral
given by an underlying local conservation law describing 
a circulatory entropy (which vanishes whenever the fluid is irrotational) 
in all even dimensions. 

Next, in section~\ref{com_class}, 
we classify kinematic and vorticity \csom/ on $n$-dimensional domains 
that move with the fluid. 
Such \csom/ arise from non-trivial local conservation laws such that 
the net flux through the moving boundary of the domain is zero. 
The main result of this classification is to show that, firstly, 
there are no vorticity \csom/ 
and, secondly, 
the only kinematic \csom/ are given by the mass integral 
and the volumetric entropy integral 
defined on any moving domain in the fluid. 
We also investigate possibilities for \csom/ defined by a conserved integral
on the boundary of a moving domain, 
in analogy with Kelvin's circulation integral in the case of 
isentropic fluid flow in $n=2$ dimensions. 
Such \csom/ correspond to conservation laws satisfying the condition of
zero net flux on the moving boundary but such that the conserved density 
in the moving domain is locally trivial 
(i.e. having the form of a spatial divergence). 
From our classification proofs for kinematic and vorticity conservation laws, 
we find that there are no moving-boundary \csom/ 
for non-isentropic compressible fluid flow in $n>1$ dimensions, 
in contrast to the generalized circulation integral which exists 
in the isentropic case for all even dimensions
as found in \cite{paperI}. 

In section~\ref{Ham_symms}, 
we then use a Hamiltonian formulation of 
the non-isentropic compressible Euler equations 
to classify all Hamiltonian symmetries corresponding to 
the kinematic and vorticity conservation laws in $n>1$ dimensions. 

In section~\ref{gas_conslaws}, 
based on the well-known relationship between 
the non-isentropic compressible Euler equations 
and the equations of gas dynamics, 
we state a complete classification of conservation laws of 
kinematic and vorticity forms for gas dynamics in $n>1$ dimensions. 
In this classification, 
the fluid \eos/ gets replaced by a state function that specifies both
the speed of sound in the gas and the thermodynamic energy density of the gas
in terms of the gas pressure and density. 
We thereby obtain conserved integrals for 
mass, momentum, energy, angular momentum and Galilean momentum, 
plus volumetric entropy and, in even dimensions, circulatory entropy,
holding for all state functions. 
The only distinguished state function for which the gas dynamics equations 
admit additional conserved integrals is shown to be the polytropic case, 
where the state function is proportional to the pressure, 
with a particular dimension-dependent proportionality coefficient 
$\gamma  =1+2/n$. 

Finally, in section~\ref{conclude}, 
we summarize our classification of fluid flow conservation laws 
in index notation
and also list the multipliers that correspond to each non-trivial conserved density. 
Proofs of the main classification theorems are given in two appendices. 

An interesting question left open for future work is 
to extend these classifications to conservation laws that depend on 
derivatives of the curl of the velocity 
or more generally on higher derivatives of the velocity itself.

\section{Non-isentropic compressible Euler Equations}
\label{kin_conslaws}

Compressible non-isentropic fluids in $\mathbb R^n$ are described 
(in the absence of external forces and viscosity) 
by a generalization of the Euler equations 
such that the equation of state for pressure becomes a function of the entropy 
in addition to the density:
\begin{align}
& 
p=P(\rho,S) ,
\label{neos}\\
&
\p_t\mtb{u}+\mtb{u}\cdot\nabla\mtb{u}+\f{1}{\rho}\nabla p=0, 
\label{nveleqn}\\
&
\p_t\rho+\nabla\cdot(\rho\mtb{u})=0. 
\label{ndeneqn}
\end{align}
Here $\mtb{u}(t,\mtb{x})$ is velocity; $\rho(t,\mtb{x})$ is density; and $S(t,\mtb{x})$ is entropy (per unit mass) of the fluid which is conserved along streamlines
\begin{equation}
\p_t S+\mtb{u}\cdot\nabla S=0. 
\label{nenteqn}
\end{equation}
In particular when $S=$const.\ or $\p P/\p S = 0$, 
the system (\ref{neos})--(\ref{ndeneqn}) 
reduces to the case of isentropic fluid flow studied in \cite{paperI},
where
\begin{equation}
p=P(\rho) ;
\label{isentropic}
\end{equation}
thus the non-isentropic case is characterized by having 
$\p P/\p S\not\equiv 0$ 
with $S$ being a dynamical variable which is non-constant across streamlines.  

We will study local conservation laws of the Eulerian system (\ref{neos})--(\ref{nenteqn}). 
Similar to the isentropic case, 
conservation laws of non-isentropic fluid flow are described by 
a local continuity equation 
\begin{equation}
D_t T+D_{\mtb{x}}\cdot\mtb{X}=0
\label{nconlaw}
\end{equation}
holding formally for all solutions of (\ref {neos})--(\ref{nenteqn}), 
where the conserved density $T$ and spatial flux $\mtb{X}$ 
are some functions of 
$t,\mtb{x},\mtb{u},\rho,S$, and $\mtb{x}$-derivatives of $\mtb{u},\rho,S$.
Here $D_t$ and $D_\mtb{x}$ denote total time and space derivatives respectively. 
In integral form, 
the continuity equation (\ref {nconlaw}) is equivalently given by
\begin{equation}
\f{d}{dt}\int_V T d^nx = -\int_{\p V}\mtb{X}\cdot\hat{\mtb{n}} d^{n-1}\sigma
\label{nintegral}
\end{equation}
where $V$ is any spatial domain in $\mathbb R^n$ 
through which the fluid is flowing, 
$\hat{\mtb{n}}$ is the outward unit normal 
and $d^{n-1}\sigma$ denotes the surface element 
of the domain boundary $\p V$.
Alternatively, 
conservation laws (\ref {nconlaw}) and (\ref{nintegral}) can be formulated 
by considering a spatial domain $V(t)$ that moves with the fluid.
Then the spatial flux through the moving boundary $\p V(t)$ is
${\boldsymbol\xi}=\mtb{X}-T\mtb{u}$ 
which is related to the conserved density $T$ by the transport equation
\begin{equation}
D_t T+\mtb{u}\cdot D_\mtb{x}T=-(\nabla\cdot\mtb{u})T-D_\mtb{x}\cdot{\boldsymbol\xi}
\label{transeqn}
\end{equation}
where $D_t +\mtb{u}\cdot D_\mtb{x}$ 
represents the total convective (material) derivative 
and $\nabla\cdot\mtb{u}$ represents 
the expansion or contraction of an infinitesimal volume moving with the fluid. 
The corresponding integral form of a fluid conservation law in a moving domain 
is thereby expressed as
\begin{equation}
\f{d}{dt}\int_{V(t)} T d^nx 
= -\int_{\p V(t)}{\boldsymbol\xi}\cdot\hat{\mtb{n}} d^{n-1}\sigma
\label{nmoveqn}
\end{equation}
in terms of the moving-flux ${\boldsymbol\xi}$ 
through the domain boundary $\p V(t)$.

For determining conserved densities $T$, 
we note the local continuity equation (\ref{nconlaw}) shows that 
${\mathcal D}_t T$ must have the form of a spatial divergence 
$D_{\mtb{x}}\cdot(-\mtb{X})$, 
where ${\mathcal D}_t$ is the total time derivative 
evaluated on solutions of (\ref {neos})--(\ref {nenteqn}). 
Hence, 
necessary and sufficient equations \cite{BCA}
for finding conserved densities $T$ 
are given by
\begin{equation}
E_\mtb{u}({\mathcal D}_t T)=E_\rho({\mathcal D}_t T)=E_S({\mathcal D}_t T)=0
\label{ndeteqn}
\end{equation}
where $E_\mtb{u}$, $E_{\rho}$, $E_S$ 
are spatial Euler operators with respect to $\mtb{u}$, $\rho$, $S$. 
To-date, a complete classification of all conserved densities 
$T(t,\mtb{x},\mtb{u},\rho,S,\nabla\mtb{u},\nabla\rho,\nabla S,\ldots )$ 
has not appeared in the literature.

Throughout, on $\mathbb R^n$, 
$\nabla$ will denote the gradient operator, 
while $\cdot$ and $\wedge$ will respectively denote the Euclidean inner product
and exterior (antisymmetric) product.

\subsection {Classification of kinematic conservation laws}

For $n$-dimensional compressible fluid flow, 
we first write down the well-known kinematic conservation laws 
\cite{Ibr,paperI}
for mass, momentum, angular momentum, Galilean momentum, and energy 
in terms of the fluid velocity $\mtb{u}$, density $\rho$ and pressure $p$, 
on any spatial domain $V(t)\subset\mathbb R^n$ transported in the fluid: 
\begin{align}
&
\f{d}{dt}\int_{V(t)} \rho d^n x = 0, 
\label{mass} \\
&
\f{d}{dt}\int_{V(t)} \rho\mtb{u} d^nx = -\int_{\p V(t)}p\hat{\mtb{n}}d^{n-1}\sigma, 
\label{mom} \\
&
\f{d}{dt}\int_{V(t)} \rho\mtb{u}\wedge\mtb{x} d^nx 
=\int_{\p V(t)}p\mtb{x}\wedge\hat{\mtb{n}}d^{n-1}\sigma, 
\label{anglmom} \\
&
\f{d}{dt}\int_{V(t)} \rho(t\mtb{u}-\mtb{x})d^nx 
= -\int_{\p V(t)}t p\hat{\mtb{n}}d^{n-1}\sigma, 
\label{Galmom} \\
&
\f{d}{dt}\int_{V(t)}\rho(\tf{1}{2}|\mtb{u}|^2+e)d^n x 
= -\int_{\p V(t)}p\mtb{u}\cdot\hat{\mtb{n}}d^{n-1}\sigma .
\label{ener}
\end{align}
Here $e$ is the internal (thermodynamic) energy density defined in terms of 
$p$ and $\rho$ by 
\begin{equation}
\frac{\p e}{\p \rho}\Bigl|_{\displaystyle S={\rm const.}} = \rho^{-2}p. 
\label{erp}
\end{equation}
We note that equation (\ref{erp}) is equivalent to the familiar
thermodynamic relation \cite{LanLif,Whi}
$d(e+p/\rho) = (1/\rho)dp$
holding for any adiabatic infinitesimal change $dS=0$
in the state of the fluid, where $e+p/\rho$ is the enthalpy. 
 
Each of these conservation laws can be readily verified to hold using 
just the dynamical Euler equations (\ref {nveleqn})--(\ref {ndeneqn}) 
for $\mtb{u}$ and $\rho$, 
plus the transport equation (\ref{nenteqn}) for $S$
in the case of the energy conservation law.  
As a consequence, 
the form of the conserved mass, momentum, angular momentum, 
and Galilean momentum 
is independent of the equation of state for $p$, 
and hence the conserved integrals 
(\ref{mass}), (\ref{mom}), (\ref{anglmom}), (\ref{Galmom}) 
are valid for non-isentropic fluid flow. 
In contrast, 
the form of the conserved energy (\ref {ener}) 
explicitly depends on the \eos/ for $p$ in terms of both $\rho$ and $S$
through the relation (\ref{erp}) for the internal energy density. 
In particular, 
for fluid flow with a non-isentropic \eos/ (\ref{neos}), 
the internal energy density is given by  
\begin{equation}
e(\rho,S)=\int\rho^{-2}P(\rho,S) d\rho. 
\label{thermoenergy}
\end{equation}
Note this defines $e$ only up to an integration constant 
given by an arbitrary function of $S$. 
Such a function $f(S)$ contributes a term of the form $\rho f(S)$ 
to the energy density $E=\rho(\f{1}{2}|\mtb{u}|^2+ e)$ of the fluid, 
and as a result, 
conservation of energy yields the conservation law
\begin{equation}
\f{d}{dt} \int_{V(t)} \rho f(S) d^n x = 0.
\label{entr}
\end{equation}
In the specific case $f(S)=S$, 
this conservation law (\ref{entr}) states that the volumetric entropy 
is conserved in any spatial domain transported by the fluid,
\begin{equation}
\f{d}{dt} \int_{V(t)} \rho S d^n x = 0 
\label{volentr}
\end{equation}
which is easy to check directly from 
the dynamical equations (\ref{ndeneqn})--(\ref{nenteqn}) for $\rho$ and $S$. 
The general case of the conservation law (\ref{entr}) can be understood 
to arise from conservation of volumetric entropy (\ref{volentr}) 
by the freedom to functionally redefine the entropy $S$ to $\tilde S= f(S)$, 
preserving the form of the entropy transport equation (\ref{nenteqn}). 

The only other kinematic conservation laws known for compressible fluid flow 
in $n>1$ dimensions are the following generalized energies
\cite{Ibr,paperI}: 
\begin{align}
&
\f{d}{dt}\int_{V(t)} (tE-\tf{1}{2}\rho\mtb{u}\cdot\mtb{x})d^nx 
= -\int_{\p V(t)}p(t\mtb{u}-\tf{1}{2}\mtb{x})\cdot\hat{\mtb{n}} d^{n-1}\sigma, 
\label{similener} \\
&
\f{d}{dt}\int_{V(t)} (t^2 E-t\rho\mtb{u}\cdot\mtb{x}+\tf{1}{2}\rho|\mtb{x}|^2)d^n x 
= -\int_{\p V(t)}pt(t\mtb{u}-\mtb{x})\cdot\hat{\mtb{n}} d^{n-1}\sigma, 
\label{dilener}
\end{align}
each holding for a polytropic \eos/ 
\begin{equation} 
p=\kappa \rho^{1+2/n}, \quad \kappa = {\rm const.}
\label{special}
\end{equation}
where 
\begin{equation}
E=\rho(\tf{1}{2}|\mtb{u}|^2+e) = \tf{1}{2}\rho|\mtb{u}|^2+\tf{1}{2}np
\label{polyenergy}
\end{equation}
is the polytropic energy density.

We now settle the natural questions of 
whether these generalized energies (\ref{similener})--(\ref{dilener}) 
hold for any non-isentropic \esos/, 
and whether the non-isentropic compressible fluid equations (\ref{neos})--(\ref{nenteqn}) 
admit any additional conservation laws of kinematic form 
\begin{equation}
T(t,\mtb{x},\mtb{u},\rho,S). 
\label{nmech}
\end{equation}

\begin{theorem}\label{kin_class}
(i) For compressible fluid flow 
with a general non-isentropic \eos/ (\ref{neos}),
the admitted kinematic conservation laws (\ref{nmech})
in any dimension $n>1$ comprise a linear combination of 
mass (\ref{mass}), momentum (\ref{mom}), angular momentum (\ref{anglmom}), 
Galilean momentum (\ref{Galmom}), energy (\ref{ener})--(\ref{thermoenergy}), 
and generalized entropy (\ref{entr}). 
(ii) Modulo a constant shift in the pressure, 
the only \esos/ for which additional conservation laws (\ref{nmech}) arise 
for non-isentropic compressible fluid flow 
is the polytropic case 
\begin{equation} 
p=\kappa(S) \rho^{1+2/n} 
\label{polytropic}
\end{equation}
given in terms of an arbitrary function of the entropy, $\kappa(S)$. 
The admitted conservation laws consist of 
the similarity energy (\ref{similener}) 
and the Galilean energy (\ref{dilener}) 
where $e(\rho,S) = \tf{1}{2}n\kappa(S) \rho^{2/n}$ 
is the internal energy density 
and $E=\rho(\f{1}{2}|\mtb{u}|^2+e) = \tf{1}{2}\rho|\mtb{u}|^2+\tf{1}{2}np$ 
is the polytropic energy density. 
\end{theorem}

We give the proof of this classification theorem in Appendix~\ref{kin_proof}.

\section{Vorticity conservation laws}
\label{vort_conslaws}

We will start by examining 
conservation of the odd-dimensional helicity integral 
and the even-dimensional enstrophy integral 
for compressible fluid flow in $n>1$ spatial dimensions. 
These integrals involve the spatial orientation tensor ${\boldsymbol\epsilon}$ 
contracted into products of the curl of the fluid velocity \cite{paperI}
\begin{equation}
{\boldsymbol\omega}=\nabla\wedge\mathbf{u}. 
\label {curl}
\end{equation} 
The dynamical equation for ${\boldsymbol\omega}$ is given by 
the curl of the Euler equation (\ref{nveleqn}) for $\mtb{u}$, 
combined with the identity 
$\mathbf{u}\cdot{\boldsymbol\omega} = 
\mathbf{u}\cdot\nabla \mathbf{u}-\tf{1}{2}\nabla(\mathbf{u}\cdot\mathbf{u})$,
which yields  
\begin{equation}
\p_t{\boldsymbol\omega} = 
\nabla\wedge ({\boldsymbol\omega}\cdot\mathbf{u}) + {\boldsymbol\sigma} ,\quad
\nabla\wedge {\boldsymbol\omega} =0 ,
\label{omegat}
\end{equation}
where 
\begin{equation}
{\boldsymbol\sigma} = \nabla \rho^{-1} \wedge \nabla p, \quad 
\nabla \wedge {\boldsymbol\sigma} = 0.
\label{sigma}
\end{equation}
We note that in the case of an isentropic \eos/ (\ref{isentropic}), 
this antisymmetric tensor (\ref{sigma}) vanishes identically, 
since $\nabla p = P'(\rho)\nabla\rho$ is proportional to $\nabla\rho$. 

In odd dimensions $n=2m+1\geq 3$, the integral \cite{ArnKhe}
\begin{equation}
\int_{V(t)} \mtb{u}\cdot{\boldsymbol\varpi}d^nx 
\label{helicityV}
\end{equation} 
defines the fluid helicity on any spatial domain $V(t) \subset\mathbb R^n$
transported in the fluid, 
where 
\begin{equation}
{\boldsymbol\varpi} = 
{\boldsymbol\epsilon}\cdot(\underbrace{
(\nabla\wedge\mathbf{u})\wedge\cdots\wedge(\nabla\wedge\mathbf{u})
}_{\textrm{$(n-1)/2$  times}}) 
= \ast({\boldsymbol\omega^m})
\label{odd}
\end{equation}
denotes the vorticity vector of the fluid.
(Here $\ast$ is the Hodge dual operator 
applied to the rank $n-1 = 2m$ skew-symmetric tensor ${\boldsymbol\omega}^m$.)
Since ${\boldsymbol\omega}$ is curl-free, 
and the tensor ${\boldsymbol\epsilon}$ is constant and skew-symmetric, 
the vorticity vector (\ref{odd}) is divergence-free 
and obeys the transport equation 
%\begin{eqnarray}
\begin{equation}
\p_t{\boldsymbol\varpi} = 
-\nabla\cdot({\boldsymbol\varpi}\wedge\mtb{u})
+m{\boldsymbol\sigma}\cdot\mtb{W}, \quad 
\mtb{W}= \ast({\boldsymbol\omega}^{m-1}), 
\label{oddtransport} 
\end{equation}
with
\begin{equation}
\nabla\cdot{\boldsymbol\varpi} = \ast(\nabla\wedge{\boldsymbol\omega^m})
=m(\nabla\wedge{\boldsymbol\omega})\cdot\ast({\boldsymbol\omega^{m-1}})
=0
\end{equation}
%\end{eqnarray}
obtained from (\ref{omegat}) and (\ref{odd}). 
Now for evaluating the time derivative of 
the helicity integral (\ref{helicityV}), 
we first use equations (\ref{oddtransport}) and (\ref{nveleqn}) to derive 
\begin{equation}
\p_t (\mtb{u}\cdot{\boldsymbol\varpi}) = 
\nabla\cdot( -\mtb{u}(\mtb{u}\cdot{\boldsymbol\varpi})
+\tf{1}{2}{\boldsymbol\varpi}|\mtb{u}|^2 ) 
-\rho^{-1}{\boldsymbol\varpi}\cdot \nabla p 
- m\mtb{W}\cdot(\mtb{u}\wedge{\boldsymbol\sigma}). 
\label{eqA}
\end{equation}
We next rewrite the pressure gradient term in (\ref{eqA}) 
by means of the identities 
$\mtb{W}\cdot(\nabla\wedge\mtb{u}) = \mtb{W}\cdot{\boldsymbol\omega}
={\boldsymbol\varpi}$
and 
$\nabla\cdot \mtb{W}=(m-1)(\nabla\wedge{\boldsymbol\omega})\cdot\ast({\boldsymbol\omega}^{m-2})=0$, 
yielding 
\begin{equation}
\rho^{-1}{\boldsymbol\varpi}\cdot\nabla p 
= \nabla\cdot(\mtb{W}\cdot(\mtb{u}\wedge \nabla p)\rho^{-1})
+ \mtb{W}\cdot(\mtb{u}\wedge{\boldsymbol\sigma}) 
\label{eqp}
\end{equation}
via integration by parts. 
We thus obtain 
\begin{equation}
\p_t (\mtb{u}\cdot{\boldsymbol\varpi}) 
+ \nabla\cdot( \mtb{u}(\mtb{u}\cdot{\boldsymbol\varpi})
-\tf{1}{2}{\boldsymbol\varpi}|\mtb{u}|^2 
+ \mtb{W}\cdot(\mtb{u}\wedge \nabla p)\rho^{-1} )
= -(1+m)\mtb{W}\cdot(\mtb{u}\wedge{\boldsymbol\sigma})
\label{oddeq}
\end{equation}
which has the form of a local continuity equation (\ref{nconlaw})
up to the term proportional to ${\boldsymbol\sigma}$ on the right-hand side.
This verifies conservation of helicity in the case of 
isentropic compressible fluid flow.
For the general case of non-isentropic compressible fluid flow, 
helicity will be conserved if and only if, for a given \eos/,
the non-vanishing ${\boldsymbol\sigma}$ term on the right-hand side 
in equation (\ref{oddeq})
reduces to a total spatial divergence. 
Necessary and sufficient conditions for this to occur are that 
the spatial Euler operators $E_\rho$, $E_S$, $E_\mtb{u}$ annihilate
the ${\boldsymbol\sigma}$ term. 

To proceed, for a general non-isentropic \eos/ $p=P(\rho,S)$, 
we note $\nabla p= P_\rho\nabla\rho + P_S\nabla S$ implies 
${\boldsymbol\sigma} = -\rho^{-2} P_S\nabla\rho \wedge \nabla S
= -\nabla e_S \wedge \nabla S$
with $e_S\not\equiv 0$,
where $e$ is the internal (thermodynamic) energy density (\ref{thermoenergy})
defined in terms of $P(\rho,S)$. 
Hence, the antisymmetric tensor (\ref{sigma}) in the non-isentropic case
is given by the curl 
\begin{equation}
{\boldsymbol\sigma} = -\nabla\wedge( e_S \nabla S ) . 
\label{sigma_curl}
\end{equation} 
This allows us to rewrite the right-hand side in equation (\ref{oddeq})
\begin{equation}
\mtb{W}\cdot(\mtb{u}\wedge{\boldsymbol\sigma}) = 
\nabla \cdot( \mtb{W}\cdot(\mtb{u}\wedge\nabla S)e_S )
- e_S\nabla\cdot({\boldsymbol\varpi}S)
\end{equation} 
via integration by parts. 
Similarly, we can rewrite the divergence term 
\begin{equation}
\nabla\cdot( \mtb{W}\cdot(\mtb{u}\wedge \nabla p)\rho^{-1} )
= \nabla\cdot( {\boldsymbol\varpi}(\rho^{-1} p + e) 
- e_S \mtb{W}\cdot(\mtb{u}\wedge\nabla S) )
\end{equation} 
by means of the identity
\begin{equation}
\nabla e = e_\rho\nabla\rho + e_S\nabla S 
= -p\nabla \rho^{-1} + e_S\nabla S = 
e_S\nabla S + \rho^{-1}\nabla p -\nabla(\rho^{-1} p)  
\end{equation} 
which follows from the thermodynamic relation (\ref{erp}). 
The conservation equation (\ref{oddeq}) 
thereby reduces to  
\begin{equation}
D_t T+D_\mtb{x}\cdot(\mtb{u}T) + D_\mtb{x}\cdot{\boldsymbol\xi} 
= (1+m) e_S\nabla\cdot({\boldsymbol\varpi}S) 
\label{helicityeq}
\end{equation} 
with 
\begin{equation}
T = \mtb{u}\cdot{\boldsymbol\varpi}, \quad 
{\boldsymbol\xi} = 
(\rho^{-1}p+e-\tf{1}{2}|\mtb{u}|^2){\boldsymbol\varpi} 
+ m e_S \mtb{W}\cdot(\nabla S\wedge\mtb{u}) .
\label{helicity}
\end{equation} 
We now observe the right-hand side in equation (\ref{helicityeq}) 
fails to be a spatial divergence when the fluid flow is non-isentropic,
since $P_S\not\equiv 0$ implies 
$E_\rho(e_S\nabla\cdot({\boldsymbol\varpi}S)) 
= e_{\rho S}\nabla\cdot({\boldsymbol\varpi}S) 
= \rho^{-2}P_S {\boldsymbol\varpi}\cdot\nabla S \not\equiv 0$. 
As a result, 
%on any spatial domain $V(t)$ transported in a compressible fluid in odd dimensions $n=2m+1$, 
the helicity integral (\ref{helicityV}) obeys 
\begin{align}
\f{d}{dt}\int_{V(t)} \mtb{u}\cdot{\boldsymbol\varpi}d^nx =& 
\int_{\p V(t)} (\tf{1}{2}|\mtb{u}|^2 -\rho^{-1}p-e){\boldsymbol\varpi}\cdot 
\hat{\mtb{n}} d^{n-1}\sigma 
+ m \int_{\p V(t)} e_S \mtb{W}\cdot(\nabla S\wedge\mtb{u}\wedge \hat{\mtb{n}})d^{n-1}\sigma 
\nonumber \\ & 
+ (1+m)\int_{V(t)} e_S\nabla\cdot({\boldsymbol\varpi}S)d^nx  
\label{eqE}
\end{align} 
where the volume term depending on the function $e_S$ 
is not conserved for any non-isentropic \eos/ (\ref{neos}). 

In even dimensions $n=2m\geq 2$, 
\begin{equation}
\int_{V(t)} \rho f(\varpi/\rho)d^nx 
\label{enstrophyV}
\end{equation}
defines the enstrophy integral \cite{ArnKhe}
in terms of an arbitrary nonlinear function $f$ of $\varpi/\rho$, 
with  
\begin{equation}
\varpi =
{\boldsymbol\epsilon}\cdot(\underbrace{
(\nabla\wedge\mathbf{u})\wedge\cdots\wedge(\nabla\wedge\mathbf{u})
}_{\textrm{$n/2$  times}}) 
=  \ast({\boldsymbol\omega}^{m})
\label{even}
\end{equation} 
denoting the vorticity scalar of the fluid. 
(In the case when $f$ is a linear function of $\varpi/\rho$, 
the integral (\ref{enstrophyV}) reduces to 
the trivially conserved circulation integral, 
which will be discussed further in the next section.) 
The transport equation obeyed by $\varpi$ is given by 
\begin{equation} 
\p_t\varpi = -\nabla\cdot(\varpi\mtb{u})- m{\boldsymbol\sigma}\cdot\mtb{w}, \quad \mtb{w}= \ast({\boldsymbol\omega}^{m-1})
\label{eventransport} 
\end{equation} 
as obtained from (\ref{even}) and (\ref{omegat}). 
To now evaluate the time derivative of the enstrophy integral (\ref{enstrophyV}), 
we start by combining the transport equation (\ref{eventransport}) 
and the Euler equation (\ref{ndeneqn}) to get 
\begin{equation} 
\p_t\widetilde\varpi = 
-\mtb{u}\cdot\nabla\widetilde\varpi - m\rho^{-1}\mtb{w}\cdot{\boldsymbol\sigma}, \quad 
\widetilde\varpi = \varpi/\rho, 
\end{equation}
and hence 
\begin{equation} 
\p_t(\rho f(\widetilde\varpi)) +\nabla\cdot(\mtb{u}\rho f(\widetilde\varpi)) 
= m f'(\widetilde\varpi)\mtb{w}\cdot{\boldsymbol\sigma} . 
\label{eveneq} 
\end{equation} 
Thus, in the case of isentropic compressible fluid flow, 
conservation of enstrophy holds due to ${\boldsymbol\sigma}=0$. 
In the general case of non-isentropic compressible fluid flow, 
we can rewrite the right-hand side in equation (\ref{eveneq}) via
\begin{equation}
\mtb{w}\cdot{\boldsymbol\sigma} = -\nabla\cdot((\mtb{w}\cdot\nabla S)e_S) 
\label{evenidentity}
\end{equation} 
which arises from the curl expression (\ref{sigma_curl})
for the antisymmetric tensor ${\boldsymbol\sigma}$
and from the divergence identity 
\begin{equation}
\nabla\cdot\mtb{w}=\ast(\nabla\wedge{\boldsymbol\omega}^{m-1})
=(m-1)(\nabla\wedge{\boldsymbol\omega})\cdot\ast({\boldsymbol\omega}^{m-2}) 
=0.
\label{dividentity}
\end{equation}
We thereby obtain, 
after integration by parts, 
\begin{equation}
{\mathcal D}_t T + \nabla\cdot(\mtb{u}T) +D_{\mtb{x}}\cdot {\boldsymbol\xi} 
= -m e_S \nabla \cdot(\mtb{w}\cdot\nabla S f'(\widetilde\varpi)) 
\label{enstrophyeq}
\end{equation} 
with 
\begin{equation}
T = \rho f(\widetilde\varpi), \quad 
{\boldsymbol\xi} = -m e_S f'(\widetilde\varpi)\mtb{w}\cdot\nabla S , 
\label{eqF}
\end{equation} 
where $f$ is a nonlinear function of $\widetilde\varpi$. 
Now, we find 
$E_\rho(e_S \nabla \cdot( (\mtb{w}\cdot\nabla S) f'(\widetilde\varpi)))
= \rho^{-3}f''(\widetilde\varpi) P_S \mtb{w}\cdot(\nabla \varpi\wedge\nabla S)
\not\equiv 0$
due to $P_S\not\equiv 0$, 
and hence the right-hand side in equation (\ref{enstrophyeq}) 
fails to be a spatial divergence
since it is not annihilated by one of the spatial Euler operators. 
Consequently, 
for fluid flow with a non-isentropic \eos/, 
the enstrophy integral (\ref{enstrophyV}) 
on any spatial domain $V(t)$ transported in the fluid
obeys 
\begin{align}
\f{d}{dt}\int_{V(t)} \rho f(\varpi/\rho)d^nx =& 
-m\int_{\p V(t)} e_S f'(\varpi/\rho)\mtb{w}\cdot (\nabla S\wedge\hat{\mtb{n}})d^{n-1}\sigma 
\nonumber\\&
-m \int_{V(t)} e_S \nabla\cdot( (\mtb{w}\cdot\nabla S) f'(\varpi/\rho))d^n x  
\end{align}  
where the volume term that depends on the function $e_S$ is not conserved. 

Essentially, the underlying reason for non-conservation of both 
the enstrophy integral (\ref{enstrophyV}) and the helicity integral (\ref{helicityV}) 
for all non-isentropic \esos/ is that $e_S = \int \rho^{-2}P_S d\rho$ 
is necessarily a non-constant function of $\rho$. 

We now address the open questions of 
whether any generalizations of the helicity (\ref{helicityV}) 
and enstrophy (\ref{enstrophyV})
are conserved for a non-isentropic \eos/ 
or, more generally, 
whether the non-isentropic fluid equations (\ref{neos})--(\ref{nenteqn}) 
admit any conservation laws of vorticity form 
\begin{equation}
T(\rho,S,\mtb{u},\ast({\boldsymbol\omega^m})) ,\quad m\geq 1
\label{nvort}
\end{equation}
where 
\begin{equation}
\ast({\boldsymbol\omega^m}) = 
\begin{cases}
\varpi &\text{if $n=2m$}\\
{\boldsymbol\varpi} &\text{if $n=2m+1$}
\end{cases}
\label{vorticity}
\end{equation}
is the vorticity scalar in even dimensions
and the vorticity vector in odd dimensions, respectively. 

\begin{theorem}\label{vort_class}
For any non-isentropic \eos/ (\ref{neos}) 
the only non-trivial vorticity conservation laws (\ref {nvort}) 
admitted for compressible fluid flow in dimensions $n>1$ are given by 
\begin{equation}
\f{d}{dt}\int_{V(t)}f(S)\varpi d^n x 
= \int_{\p V(t)} e_S \mtb{w} \cdot( \nabla F(S) \wedge \hat{\mtb{n}} ) d^{n-1}\sigma
\label{vortconslaw}
\end{equation} 
for even dimensions $n=2m$, 
where $f(S) = F'(S)$ is a non-constant function of the entropy $S$. 
In particular, 
there are no special \esos/ for which additional vorticity conservation laws
are admitted in any even dimension, 
and no non-trivial vorticity conservation laws are admitted 
in any odd dimension $n>1$. 
\end{theorem}

The proof of this classification theorem will be given in Appendix~\ref{vort_proof}. 

The even-dimensional vorticity conservation law (\ref{vortconslaw})
can be rewritten through integration by parts via the identities
\begin{equation}
\varpi = \mtb{w}\cdot {\boldsymbol\omega} 
= \mtb{w}\cdot(\nabla \wedge \mtb{u}), \quad
\nabla\cdot \mtb{w} =0 . 
\label{evenvortids}
\end{equation}
This yields the equivalent conserved integral 
\begin{equation}
\f{d}{dt}\int_{V(t)} \mtb{w} \cdot( \mtb{u}\wedge\nabla f(S) ) d^n x
= \int_{\p V(t)} \big( \tf{1}{2}|\mtb{u}|^2 \nabla f(S) - e_S \nabla F(S) \big) 
\cdot(\mtb{w} \cdot\hat{\mtb{n}} ) d^{n-1}\sigma
\label{circconslaw}
\end{equation} 
where $n=2m$. 
We then see that the case $f(S)=S$ has the form of a circulatory entropy
\begin{equation}
\f{d}{dt}\int_{V(t)} \mtb{w} \cdot( \mtb{u}\wedge\nabla S ) d^n x
= \int_{\p V(t)} \big( \tf{1}{2}|\mtb{u}|^2 - e_S S \big) 
\mtb{w}\cdot(\nabla S\wedge\hat{\mtb{n}} ) d^{n-1}\sigma
\label{circentr}
\end{equation} 
which vanishes whenever the fluid flow is either irrotational, i.e. 
${\boldsymbol\omega} = 0$ and thus 
$\mtb{w}= \ast({\boldsymbol\omega^{m-1}}) = 0$,
or isentropic, i.e. $S=$const. and thus $\nabla S=0$. 
Hence, in the general case, the conserved integral (\ref{circconslaw})
can be understood to arise from conservation of circulatory entropy (\ref{circentr})
by the freedom to functionally redefine the entropy $S$ to $\tilde S = f(S)$. 

To conclude, we point out that the respective transport equations
for the vorticity (\ref{vorticity}) in even and odd dimensions
each yield locally trivial conservation laws
(therefore falling outside of Theorem~\ref{vort_class})
in which 
\begin{equation}
T=\ast({\boldsymbol\omega}^m)
=\nabla \cdot \ast(\mathbf{u}\wedge {\boldsymbol\omega}^{m-1}) 
= {\rm div}\mathbf{\Theta}
\label{trivial}
\end{equation}
is a spatial divergence.
In particular, 
in even dimensions $n=2m$,
we see the transport equation (\ref{eventransport}) for $\varpi$
takes the form of a conservation law (\ref{transeqn}) given by 
the conserved density 
$T=\varpi=\nabla\cdot(\mtb{w}\cdot\mtb{u})$,
with the moving spatial flux 
${\boldsymbol \xi}=-m e_S \mtb{w}\cdot\nabla S$
obtained from the identity (\ref{evenidentity}). 
We note this is precisely the local conservation law that underlies 
the conserved integral (\ref{vortconslaw}) when the fluid flow is isentropic. 
Similarly, in odd dimensions $n=2m+1$ ($m\geq 1$),
by combining the transport equation (\ref{oddtransport}) for ${\boldsymbol\varpi}$
with the curl expression (\ref{sigma_curl}),
we obtain a conservation law (\ref{transeqn}) where
$T={\boldsymbol\varpi} =\nabla\cdot(\mtb{W}\cdot\mtb{u})$ 
is a conserved vector-density and 
${\boldsymbol \xi}=-\mtb{u}\otimes{\boldsymbol\varpi} 
-m e_S \mtb{W}\cdot\nabla S$
is the moving spatial tensor-flux.

\section{Constants of motion}
\label{com_class}

A fluid conservation law in integral form (\ref{nmoveqn})
on a domain $V(t)\subset\mathbb R^n$ moving with the fluid 
yields a \com/
\begin{equation}
\f{d}{dt}\int_{V(t)} T d^nx =0
\label{com}
\end{equation}
if the net flux through the moving domain boundary $\p V(t)$ is zero
for all solutions of the Eulerian fluid equations 
(\ref{nveleqn})--(\ref{nenteqn}). 
This condition has an equivalent formulation as a local continuity equation
in the form (\ref{transeqn}) such that the moving flux vector 
${\boldsymbol\xi}$ is divergence free in the domain $V(t)$.
Thus, 
a conserved density $T$ determines a \com/ (\ref{com}) if and only if 
\begin{equation}
{\mathcal D}_t T+ D_{\mtb{x}}\cdot(\mtb{u} T)= 0 . 
\label{comconslaw}
\end{equation}
The resulting conserved integral (\ref{com}) 
will be called a {\em moving-domain \com/} 
provided the conserved density is non-trivial, i.e.
if $T$ is not a spatial divergence ${\rm div}\boldsymbol{\Theta}$
such that the vector $\boldsymbol{\Theta}$ is a local function of
$t,\mtb{x},\mtb{u},\rho,S$, and $\mtb{x}$-derivatives of $\mtb{u},\rho,S$.

Our classification of non-trivial kinematic and vorticity conservation laws
in Theorems~\ref{kin_class} and~\ref{vort_class} 
provides an immediate corresponding classification of
moving-domain \csom/ of kinematic form (\ref{nmech}) 
and vorticity form (\ref{nvort})--(\ref{vorticity}). 

\begin{proposition}\label{kin_com}
For compressible fluid flow in $n>1$ dimensions, 
the only kinematic \csom/ 
\begin{equation}
\f{d}{dt} \int_{V(t)} T(t,\mtb{x},\mtb{u},\rho,S) d^nx =0
\end{equation}
on an arbitrary moving domain $V(t)$ transported by the fluid 
consist of mass
\begin{equation}
%\f{d}{dt}
\int_{V(t)} \rho d^n x 
%= 0
\end{equation}
and generalized entropy
\begin{equation}
%\f{d}{dt} 
\int_{V(t)} \rho f(S) d^n x 
%= 0
\end{equation}
where $f$ is any non-constant function of $S$.
These \csom/ hold for all \esos/ (\ref{neos}).
\end{proposition}

In particular, 
no additional kinematic \csom/ arise 
for any special (non-isentropic or isentropic) \eos/ 
or in any special dimension. 

\begin{proposition}\label{vort_com}
There are no vorticity \csom/ 
\begin{equation}
\f{d}{dt}\int_{V(t)} T(\rho,S,\mtb{u},\varpi) d^nx =0\ 
\text{if}\ n=2m 
\quad\text{or}\quad 
\f{d}{dt}\int_{V(t)} T(\rho,S,\mtb{u},{\boldsymbol\varpi}) d^nx =0\ 
\text{if}\ n=2m+1
\end{equation} 
for compressible fluid flow with any non-isentropic \eos/ (\ref{neos})
in $n>1$ dimensions. 
In the case of isentropic \esos/ (\ref{isentropic}), 
where 
\begin{equation}
e=\int \rho^{-2}P(\rho)d\rho \quad(\text{with $e_S=0$})
\end{equation}
is the internal (thermodynamic) energy density, 
the even-dimensional enstrophy
\begin{equation}
%\f{d}{dt}
\int_{V(t)} \rho f(\varpi/\rho)d^nx
\quad (n=2m)
%= 0
\end{equation}
given by any nonlinear function $f$ 
yields the only vorticity \com/ 
on an arbitrary moving domain $V(t)$ transported by the fluid
in $n>1$ dimensions. 
\end{proposition}

In addition to the classifications given by these two Propositions, 
we can consider non-trivial \csom/ of the form 
\begin{equation}
\f{d}{dt} \int_{\p V(t)}\boldsymbol{\Theta}\cdot\hat{\mtb{n}} d^{n-1}\sigma =0
\label{mbcom}
\end{equation}
holding formally for all solutions of the fluid equations 
(\ref{nveleqn})--(\ref{nenteqn}),
where $\p V(t)$ is the moving boundary of a domain $V(t)$ 
transported by the fluid. 
Through the divergence theorem, any conserved integral (\ref{mbcom})
arises from a locally trivial conserved density 
$T ={\rm div}\boldsymbol{\Theta}$ 
satisfying the moving-flux condition (\ref{comconslaw})
in the domain $V(t)$.
In particular, 
a vector function $\boldsymbol{\Theta}$ yields 
a non-trivial conserved integral (\ref{mbcom}) if and only if
\begin{equation}
{\mathcal D}_t \boldsymbol{\Theta} + \mtb{u}(\nabla\cdot\boldsymbol{\Theta})
= D_{\mtb{x}}\cdot\boldsymbol{\Psi}
\label{mbcomconslaw}
\end{equation}
holds for some antisymmetric tensor function $\boldsymbol{\Psi}$
in terms of 
$t,\mtb{x},\mtb{u},\rho,S$, and $\mtb{x}$-derivatives of $\mtb{u},\rho,S$.
If $\boldsymbol{\Theta}$ is not identically divergence-free,
the resulting non-trivial conserved integral (\ref{mbcom}) 
will be called a {\em moving-boundary \com/}. 
We remark that the corresponding local continuity equation (\ref{mbcomconslaw})
can be viewed as a special type of lower-degree conservation law 
\cite{paperIII}. 

For isentropic fluid flow, 
a well-known example of a moving-boundary \com/ is given by 
Kelvin's circulation theorem \cite{Lig} in $n=2$ dimensions
which states the circulation of the fluid velocity 
around any closed curve transported in the fluid 
is a constant of the fluid motion.
The circulation integral has the form 
\begin{equation}
\oint_{C(t)} \mathbf{u} \cdot d\mathbf{s}
= \oint_{C(t)} \mathbf{u} \cdot\ast\hat{\mathbf{n}}d\sigma 
= \oint_{C(t)} \boldsymbol{\epsilon}\cdot
(\mathbf{u}\wedge\hat{\mathbf{n}}) d\sigma 
= -\int_{V(t)} \varpi d^2 x 
\label{kelvin}
\end{equation}
related to the two-dimensional vorticity scalar
$\varpi ={\boldsymbol\epsilon}\cdot{\boldsymbol\omega}
=\ast(\nabla\wedge\mathbf{u})$
(defined in terms of the spatial orientation tensor $\boldsymbol{\epsilon}$ 
in $\mathbb R^2$),
where 
$C(t)$ is any closed curve bounding a domain $V(t)$ 
transported by the fluid in $\mathbb R^2$, 
$\ast\hat{\mathbf{n}}$ is a unit tangent vector along $C(t)$ 
or equivalently $\hat{\mathbf{n}}$ is a unit normal vector, 
and $d\sigma$ is the arclength element for $C(t)$. 
As shown in \cite{paperI}, 
the circulation (\ref{kelvin}) has a generalization
to all even dimensions $n=2m$ $(m>1)$:
\begin{equation}
\oint_{\p V(t)}(\mathbf{u}\wedge{\boldsymbol\omega}^{m-1})\cdot d\mathbf{A}
= \int_{\p V(t)}(\mathbf{u}\wedge{\boldsymbol\omega}^{m-1})\cdot\ast\hat{\mathbf{n}}d^{2m-1}\sigma 
= -\int_{V(t)} \varpi d^{2m} x .
\label{vorticitycom}
\end{equation}
This integral defines a constant of the fluid motion 
for isentropic fluid flow, 
where $d\mathbf{A}= \ast\hat{\mathbf{n}}d^{2m-1}\sigma$ 
denotes the surface element for the moving-boundary hypersurface $\p V(t)$ 
in even dimensions $n=2m$ 
analogously to $d\mathbf{s}=\ast\hat{\mathbf{n}}d\sigma$ 
for moving-boundary curves in two dimensions.

The natural question now arises as to whether 
the transport equations (\ref{eventransport}) and (\ref{oddtransport}) 
for the vorticity (\ref{vorticity}) in any even or odd dimension $n>1$
yield a non-trivial \com/ (\ref{mbcom}) when the fluid flow is non-isentropic.
Our classification proof for vorticity conservation laws in Theorem~\ref{vort_class}
actually settles this question in a slightly more general form. 

\begin{proposition}\label{mbcom_class}
For compressible fluid flow with any non-isentropic \eos/ (\ref{neos})
in $n>1$ dimensions, 
there are no moving-boundary \csom/ (\ref{mbcom}) 
whose corresponding local continuity equation (\ref{comconslaw}) 
for $T ={\rm div}\boldsymbol{\Theta}$ is of 
vorticity form (\ref{nvort})--(\ref{vorticity}). 
In the case of isentropic \esos/ (\ref{isentropic}),
the only moving-boundary \com/ with such a form is given by 
the generalized circulation (\ref{vorticitycom})
in all even dimensions. 
\end{proposition}

In particular, 
for non-isentropic fluid flow, the vorticity transport equations
written in the moving-boundary form (\ref{trivial}) show that
\begin{equation}
\f{d}{dt} \oint_{\p V(t)} (\mathbf{u}\wedge{\boldsymbol\omega}^{m-1})\cdot d\mathbf{A}
= m \oint_{\p V(t)} e_S (\nabla S\wedge{\boldsymbol\omega}^{m-1})\cdot d\mathbf{A}
\end{equation}
in even dimensions $n=2m$,
while in odd dimensions $n=2m+1$
\begin{equation}
\f{d}{dt} \oint_{\p V(t)} (\mathbf{u}\wedge{\boldsymbol\omega}^{m-1})\cdot d\mathbf{A}
= m \oint_{\p V(t)} e_S (\nabla S\wedge{\boldsymbol\omega}^{m-1})\cdot d\mathbf{A}
- \oint_{\p V(t)} \mathbf{u}\ {\boldsymbol\omega}^m \cdot d\mathbf{A} 
\end{equation}
with $d\mathbf{A}= \ast\hat{\mathbf{n}}d^{2m-1}\sigma$, 
where $e_S\not\equiv 0$ is a non-constant function of $\rho$
for any non-isentropic \eos/ (\ref{neos}).

\section{Hamiltonian correspondence}
\label{Ham_symms}

A Hamiltonian formulation for the non-isentropic 
compressible Euler equations (\ref{neos})--(\ref{nenteqn}) in $n>1$ dimensions 
(cf \cite{Kup}) is given by 
\begin{equation}
\p_t \begin{array}({c})\mtb{u}\\\rho\\S \end{array}
=\mathcal{H}\begin{array}({c})
\delta E/\delta \mtb{u}\\ 
\delta E/\delta \rho \\ 
\delta E/\delta S\end{array} ,\quad
E=\tf{1}{2}\rho|\mtb{u}|^2 +\rho e(\rho,S) 
= \rho\big(\tf{1}{2} |\mtb{u}|^2 +\int\rho^{-2}P(\rho,S) d\rho\big) , 
\label{Hameqns}
\end{equation}
in terms of the Hamiltonian operator
\begin{equation}
\mathcal{H}=\begin{array}({ccc})
\rho^{-1}(\nabla\wedge\mtb{u})\cdot&-\nabla &\rho^{-1}\nabla S \\
-\nabla\cdot &0 &0\\
-(\rho^{-1}\nabla S)\cdot &0 &0 \end{array}
\label{Hamop}
\end{equation}
where $E$ is the energy density of the fluid. 
This operator (\ref{Hamop}) determines a Poisson bracket 
\begin{align}
\{{\mathcal F},{\mathcal G}\}_\mathcal{H} 
&=
\displaystyle \int 
\begin{array}({ccc})\delta F/\delta \mathbf{u}&\delta F/\delta \rho&\delta F/\delta S \end{array}
\mathcal{H}
\begin{array}({c})\delta G/\delta \mathbf{u}\\ \delta G/\delta \rho \\ \delta G/\delta S \end{array} 
d^n x 
\\
&= 
\displaystyle \int 
\rho^{-1}(\nabla\wedge\mathbf{u})\cdot (\delta F/\delta \mathbf{u}\wedge \delta G/\delta \mathbf{u})
+\delta G/\delta \mathbf{u}\cdot(\nabla \delta F/\delta \rho 
-(\rho^{-1}\delta F/\delta S)\nabla S )
\nonumber\\
&\qquad 
-\delta F/\delta \mathbf{u}\cdot(\nabla \delta G/\delta \rho 
-(\rho^{-1}\delta G/\delta S)\nabla S )
\; d^n x
\nonumber
\end{align} 
satisfying (modulo divergence terms) antisymmetry 
and the Jacobi identity \cite{Olv}, 
for arbitrary functionals 
${\mathcal F}=\int F d^n x$ and ${\mathcal G}=\int G d^n x$ 
where $F$ and $G$ are functions of 
$t,\mathbf{x},\mathbf{u},\rho,S$, and $\mathbf{x}$-derivatives of $\mtb{u},\rho,S$. 
Here $\delta /\delta \mathbf{u}$, $\delta /\delta \rho$, $\delta /\delta S$
denote variational derivatives, which respectively coincide with 
the spatial Euler operators $E_\mathbf{u}$, $E_\rho$, $E_S$ 
when acting on functions that do not contain time derivatives of 
$\mathbf{u},\rho,S$. 

Similarly to the isentropic case described in \cite{paperI}, 
the Hamiltonian operator $\mathcal{H}$ gives rise to an explicit mapping 
\begin{equation}
-\mathcal{H}\begin{array}({c})
\delta T/\delta \mtb{u}\\\delta T/\delta \rho\\ \delta T/\delta S \end{array}
=\widehat{\mtb{X}}\begin{array}({c})\mtb{u}\\\rho\\S \end{array}
= \begin{array}({c})\hat{\boldsymbol\eta}\\\hat\eta\\\hat\phi\end{array}
\label{Hamsymmmap}
\end{equation}
which produces infinitesimal symmetries 
$\widehat{\mtb{X}}= 
\hat{\boldsymbol\eta}\rfloor\p_\mtb{u}+\hat{\eta}\p_{\rho}+\hat{\phi}\p_S$ 
of the non-isentropic Euler equations from conserved densities $T$. 
In particular, 
these components of the symmetry generator 
\begin{equation}
\hat{\boldsymbol\eta}=
-\rho^{-1}(\nabla\wedge\mtb{u})\cdot \delta T/\delta \mtb{u}
+ \nabla\delta T/\delta \rho
-(\rho^{-1}\delta T/\delta S)\nabla S,\quad
\hat\eta=
\nabla\cdot\delta T/\delta \mtb{u},\quad
\hat\phi=
\rho^{-1}\nabla S\cdot\delta T/\delta \mtb{u}
\label{Xcomp}
\end{equation}
%\end{eqnarray}
will satisfy the infinitesimal invariance equations \cite{Olv,BCA}
\begin{equation}\label{symmdeteqns}
\begin{aligned}
& 
D_t\hat{\boldsymbol\eta}+\mtb{u}\cdot\nabla\hat{\boldsymbol\eta}
+\hat{\boldsymbol\eta}\cdot\nabla\mtb{u}
-\rho^{-2}\hat\eta \nabla P+\rho^{-1}\nabla(P_\rho\hat\eta+P_S\hat\phi)=0,
\\
&
D_t\hat\eta+\nabla\cdot\mtb{u}\hat\eta+\rho\nabla\cdot\hat{\boldsymbol\eta}=0,
\quad 
D_t\hat\phi+\hat{\boldsymbol\eta}\cdot\nabla S+ \mtb{u}\cdot\nabla \hat\phi=0
\end{aligned}
\end{equation}
for all solutions of the non-isentropic 
Euler equations (\ref {neos})--(\ref{nenteqn}). 

Evaluating the mapping (\ref{Hamsymmmap})--(\ref{Xcomp}) 
for the kinematic and vorticity conserved densities 
given by the conserved integrals 
in Theorems~\ref{kin_class} and~\ref{vort_class}, 
we obtain the following two tables of symmetries,
all of which can be written as point transformations
$\mtb{X}= \tau \p_t + {\boldsymbol\xi}\rfloor\mtb{x} 
+ \boldsymbol\eta\rfloor\p_\mtb{u}+\eta\p_{\rho}+\phi\p_S$
given by 
\begin{equation}
\hat{\boldsymbol\eta}={\boldsymbol\eta}-\tau \p_t\mathbf{u}-{\boldsymbol\xi}\cdot\nabla \mathbf{u}, \quad
\hat\eta=\eta-\tau \p_t \rho -{\boldsymbol\xi}\cdot\nabla\rho ,\quad
\hat\phi=\phi-\tau \p_t S -{\boldsymbol\xi}\cdot\nabla S ,
\end{equation}
where ${\boldsymbol\eta}$, $\eta$, $\phi$ are functions of 
$t,\mathbf{x},\rho,S,\mathbf{u}$, 
while $\tau$, ${\boldsymbol\xi}$ are functions only of $t$, $\mathbf{x}$. 

For general non-isentropic \esos/ (\ref{neos}):\\
\begin{tabular}{|c|c|c|c|}
\hline
Conserved Density $T$ & Conservation Law 
& Symmetry $\mtb{X}$ & Description \\
\hline
$\rho$ & Mass & 
0 &  Nil\\
$\rho\mtb{u}$ & Momentum &
$\p_{\mtb{x}}$ & Space translations\\
$\rho\mtb{u}\wedge\mtb{x}$ & Angular Momentum &
$\mtb{x}\wedge\p_{\mtb{x}}+\mtb{u}\wedge\p_{\mtb{u}}$ & Rotations\\
$\rho(t\mtb{u}-\mtb{x})$ & Galilean Momentum &
$t\p_{\mtb{x}}+\p_{\mtb{u}}$ & Galilean boosts\\
$\f{1}{2}\rho|\mtb{u}|^2+\rho e$ & Energy &
$\p_t$ & Time translation\\
$\rho S$ & Volumetric Entropy &
$0$ & Nil\\
$\varpi S\equiv *({\boldsymbol\omega}^{m-1}\wedge \mtb{u}\wedge\nabla S)$ 
& Circulatory Entropy ($n=2m$) & 
$0$ & Nil\\
\hline
\end{tabular}\\
Here $e$ is the internal (thermodynamic) energy density (\ref{thermoenergy}). 
Note the entropy (per unit mass) $S$ can be replaced by an arbitrary function 
$f(S)$. 

For polytropic \esos/ (\ref{polytropic}):\\
\begin{tabular}{|c|c|c|c|}
\hline
Conserved Density $T$ & Conservation Law 
& Symmetry $\mtb{X}$ & Description \\
\hline
$tE-\f{1}{2}\rho(\mtb{u}\cdot\mtb{x})$ & Similarity Energy &
%$t\p_t+\f{1}{2}\mtb{x}\rfloor\p_\mtb{x}$ &Similarity Scaling\\&&\quad $-\f{1}{2}\mtb{u}\p_\mtb{u}-\f{1}{2}n\rho \p_\rho$ &\\ 
$\begin{array}{l} t\p_t+\f{1}{2}\mtb{x}\rfloor\p_\mtb{x} \\ \quad
-\f{1}{2}\mtb{u}\p_\mtb{u}-\f{1}{2}n\rho \p_\rho \end{array}$
& Similarity Scaling \\
$t^2E-t\rho(\mtb{u}\cdot\mtb{x})+\f{1}{2}\rho|\mtb{x}|^2$ & Galilean Energy &
$\begin{array}{l} t^2\p_t+t\mtb{x}\rfloor\p_\mtb{x} \\ \quad
-(t\mtb{u}-\mtb{x})\p_\mtb{u}-nt\rho\p_\rho \end{array}$
& Galilean Dilation \\
\hline
\end{tabular}\\
Here $E=\rho(\f{1}{2}|\mtb{u}|^2+e) 
= \f{1}{2}\rho|\mtb{u}|^2+\f{1}{2}n\kappa(S) \rho^{1+2/n}$ 
is the polytropic energy density. 
%(\ref{polyenergy})

As none of these symmetry generators $\mtb{X}$ contain the pressure $p$, 
we see that the symmetry structure produced via 
the Hamiltonian mapping (\ref{Xcomp}) 
for the non-isentropic Euler equations (\ref {neos})--(\ref{nenteqn})
is the same as in the isentropic case presented in \cite{paperI}. 
This structure can be geometrically summarized as follows. 

Recall, 
the Killing equation on $\mathbb R^n$ is given by 
\begin{equation}
{\mathcal L}_{\boldsymbol\zeta}\mathbf{g} = 0
\label{killing}
\end{equation}
for a vector ${\boldsymbol\zeta}(\mathbf{x})$,
where ${\mathcal L}$ denotes the Lie derivative 
and $\mathbf{g}$ is the Euclidean metric tensor. 
Also recall, 
a vector ${\boldsymbol\zeta}(\mathbf{x})$ is irrotational on $\mathbb R^n$ 
if $\nabla \wedge {\boldsymbol\zeta} = 0$. 

\begin{proposition}\label{Ham_symm_class}
(i) 
For a general \eos/ (\ref{neos}),
the Hamiltonian symmetries corresponding to the kinematic conserved densities 
for energy (\ref{ener}), momentum (\ref{mom}), 
angular momentum (\ref{anglmom}) and Galilean momentum (\ref{Galmom})
comprise the generators of the Galilean group in $n>1$ dimensions given by 
\begin{equation}
\mtb{X}= \p_t ,
\quad
\mtb{X}= {\boldsymbol\zeta}\rfloor\p_\mathbf{x} + \tf{1}{2}\mathbf{u}\cdot(\nabla\wedge{\boldsymbol\zeta})\rfloor \p_\mathbf{u},
\quad
\mtb{X}= t{\boldsymbol\chi}\rfloor\p_\mathbf{x} + {\boldsymbol\chi}\rfloor \p_\mathbf{u},
\end{equation}
in terms of solutions ${\boldsymbol\zeta}(\mathbf{x})$ of the Killing equation,
and irrotational solutions ${\boldsymbol\chi}(\mathbf{x})$ of the Killing equation. 
(ii) 
The additional Hamiltonian symmetries, 
consisting of the similarity scaling and Galilean dilation
which correspond to the generalized energy densities 
(\ref{similener}) and (\ref{dilener}) 
in the case of a polytropic \eos/ (\ref{polytropic}), 
generate an extension of the Galilean group given by 
\begin{equation}
\begin{gathered}
\mtb{X}= \p_t ,
\quad
\mtb{X}=  \lambda t\p_t + {\boldsymbol\zeta} \rfloor\p_\mathbf{x} - \tf{1}{2}\lambda n\rho\p_\rho +\tf{1}{2}( \mathbf{u}\cdot(\nabla\wedge{\boldsymbol\zeta})
- \lambda \mathbf{u})\rfloor\p_\mathbf{u},
\\
\mtb{X}=  \tf{1}{2}\sigma t^2\p_t + t{\boldsymbol\chi} \rfloor\p_\mathbf{x}  -\tf{1}{2}\sigma nt\rho \p_\rho +({\boldsymbol\chi}-\tf{1}{2}t\sigma \mathbf{u})\rfloor \p_\mathbf{u}, 
\end{gathered}
\end{equation}
where the vector ${\boldsymbol\zeta}(\mathbf{x})$ is the solution of the homothetic Killing equation ${\mathcal L}_{\boldsymbol\zeta}\mathbf{g} = \lambda \mathbf{g}$, $\lambda =$const., 
and the vector ${\boldsymbol\chi}(\mathbf{x})$ 
is the irrotational solution of the homothetic Killing equation 
${\mathcal L}_{\boldsymbol\chi}\mathbf{g} = \sigma \mathbf{g}$, $\sigma =$const.. 
\end{proposition}

In contrast, 
the conserved densities for mass (\ref{mass}), 
generalized entropy (\ref{entr}), 
and circulatory entropy (\ref{circconslaw}) 
(or equivalently (\ref{vortconslaw}))
are mapped into the trivial symmetry $\mtb{X}=0$. 
These conserved densities comprise all of the Hamiltonian Casimirs
for the non-isentropic Euler equations (\ref {neos})--(\ref{nenteqn})
in $n>1$ dimensions.

\section{Gas Dynamics Conservation Laws}
\label{gas_conslaws}

The Euler equations (\ref {neos})--(\ref{nenteqn}) for 
non-isentropic compressible fluid flow have an equivalent formulation 
in which the \eos/ (\ref{neos}) is inverted to give entropy 
as a function of pressure and density, 
$S=S(p,\rho)$ with $\p S/\p p\not\equiv 0$.
The entropy transport equation (\ref{nenteqn}) then becomes
\begin{equation}
\p_tp+\mtb{u}\cdot\nabla p+F(p,\rho)\nabla\cdot\mtb{u}=0 
\label{npeqn}
\end{equation}
in terms of the function 
\begin{equation}
F(p,\rho)=-\rho \frac{\p S/\p \rho}{\p S/\p p}
\label{Feqn}
\end{equation}
which replaces the \eos/ (\ref{neos}).
For any specific choice of $F(p,\rho)$, 
the entropy function $S(p,\rho)$ can be recovered 
as a solution of the linear equation 
\begin{equation}
\rho S_\rho + F(p,\rho) S_p=0 .
\label{Seqn}
\end{equation}
In particular, we can identify $S(p,\rho)$ with the constant of integration
arising from the characteristic ODE
\begin{equation}
\frac{dp}{d\rho}=F(p,\rho)/\rho . 
\label{Peqn}
\end{equation}
Note that the simple state function 
\begin{equation}
F=\gamma p, \quad \gamma = {\rm const.}
\label{Fpoly}
\end{equation}
corresponds to a polytropic equation of state (\ref{polytropic})
which is equivalent to an entropy function of the form 
$S=S(p/\rho^\gamma)$. 

The resulting formulation of compressible fluid flow thus consists of 
the dynamical equations (\ref{npeqn}), (\ref{nveleqn}), (\ref{ndeneqn})
for pressure $p$, velocity $\mtb{u}$, and density $\rho$, respectively. 
These are precisely the equations of adiabatic gas dynamics \cite{Whi}. 
Physically, the state function $F(p,\rho)$ is then related to 
the sound speed $c$ in the gas by $F/\rho = c^2$ 
as seen from equation (\ref{Peqn}). 
The state function also determines the internal energy density 
$e$ of the gas from the thermodynamic relations 
$de=-pd(1/\rho)= e_\rho d\rho + e_p dp$
combined with equation (\ref{Peqn}), 
which yields 
\begin {equation}
\rho e_\rho+F(p,\rho)e_p = \f{p}{\rho} 
\label{eeqn}
\end{equation}
for the function $e(p,\rho)$. 
Note the homogeneous part of this linear equation (\ref{eeqn}) 
involves the same linear differential operator $\rho \p_\rho + F(p,\rho)\p_p$ 
appearing in the entropy equation (\ref{Seqn}), 
so thus the general homogeneous solution of equation (\ref{eeqn}) 
is given by $f(S(p,\rho))$ in terms of 
any particular solution of equation (\ref{Seqn}), 
where $f$ is an arbitrary function of $S(p,\rho)$.

Since the dynamical pressure equation (\ref{Peqn}) is related to 
the entropy transport equation (\ref{nenteqn}) by an invertible change of
variables $p=P(\rho,S) \leftrightarrow S=S(p,\rho)$ 
(i.e. a point transformation) when $S_p\not\equiv0$ or $P_S\not\equiv 0$,
there is consequently a one-to-one correspondence between 
the local conservation law structure of 
the gas dynamics equations (\ref{npeqn}), (\ref{Feqn}), 
(\ref{nveleqn}), (\ref{ndeneqn}) 
and the non-isentropic Euler equations (\ref {neos})--(\ref{nenteqn}),
as shown by general results in \cite{BTA,BCA}. 
A complete classification of kinematic and vorticity conservation laws 
for gas dynamics in $n>1$ dimensions now follows as a corollary 
to Theorems~\ref{kin_class} and~\ref{vort_class}. 

\begin{theorem}\label{gas_kin_class}
(i) For a general state function $F(p,\rho)$, 
all gas dynamics conserved densities of kinematic form 
$T(t,\mtb{x},\mtb{u},\rho,p)$
in any dimension $n>1$ are given by 
a linear combination of the conservation laws for 
mass (\ref{mass}), momentum (\ref{mom}), angular momentum (\ref{anglmom}), 
Galilean momentum (\ref{Galmom}), plus 
energy (\ref{ener}) and volumetric entropy (\ref{volentr}) 
where $e=e(p,\rho)$ is any non-homogeneous solution of 
the thermodynamic energy equation (\ref{eeqn}) 
and $S=S(p,\rho)$ is the general solution of the entropy equation (\ref{Seqn}).
(ii) The only state function $F(p,\rho)$ for which additional 
kinematic conserved densities arise is the polytropic case (\ref{Fpoly}) 
with dimension-dependent coefficient $\gamma=1+2/n$. 
The admitted conservation laws consist of 
the similarity energy (\ref{similener}) 
and the Galilean energy (\ref{dilener}) 
given in terms of the polytropic energy density (\ref{polyenergy}). 
\end{theorem}

\begin{theorem}\label{gas_vort_class}
(i) For a general state function $F(p,\rho)$, 
all gas dynamics conserved densities of vorticity form 
$T(\rho,p,\mtb{u},\varpi)$ if $n=2m$
%\quad\text{or}\quad 
or $T(\rho,p,\mtb{u},{\boldsymbol\varpi})$ if $n=2m+1$
(with $m\ge 1$) 
consist of the conservation law for circulatory entropy (\ref{circentr}) 
in even dimensions, as expressed in terms of 
$\mtb{w}=*({\boldsymbol\omega}^{m-1})$, 
$\nabla S= S_p(\nabla p-\rho^{-1}F(p,\rho)\nabla \rho)$, and $e_S=e_p/S_p$, 
where 
$S=S(p,\rho)$ is the general solution of the entropy equation (\ref{Seqn}), 
$e=e(p,\rho)$ is any non-homogeneous solution of 
the thermodynamic energy equation (\ref{eeqn}), 
and $\mtb{w}$ is an antisymmetric tensor related to 
the vorticity scalar $\varpi$ by the identities (\ref{evenvortids}). 
(ii) There are no state functions $F(p,\rho)$ for which additional 
vorticity conserved densities arise in any even or odd dimension $n>1$. 
\end{theorem}

\section{Multipliers}
\label{conclude}

In index notation, 
the Euler equations for inviscid non-isentropic compressible fluid flow 
on $\mathbb{R}^n$ in Cartesian coordinates $x^i$ ($i=1,\ldots,n$)
are given by
\begin{gather}
u^i_t + u^j u^i{}_{,j} +\rho^{-1}p_,{}^i =0, 
\label{eqn_1}\\
\rho_t +(\rho u^i)_{,i}=0, 
\label{eqn_2} \\
S_t+u^i S_{,i}=0, 
\label{eqn_3}
\end{gather}
together with the \eos/ 
\begin{equation}
p=P(\rho,S), \quad
P_S \not\equiv 0.
\label{eqn_4}
\end{equation}
Similarly to the isentropic case studied in \cite{paperI},
any local conservation law $\mathcal{D}_t T+D_i X^i=0$ that holds formally for
all solutions of these equations (\ref{eqn_1})--(\ref{eqn_4})
can be written in an equivalent {\it characteristic form}
\begin{equation}
D_t T + D_i \tilde{X}^i = 
( u^i_t+u^j u^i{}_{,j} + \rho^{-1} P_\rho\rho{}_,{}^i 
+ \rho^{-1} P_S S{}_,{}^i ) Q_i
+ (\rho_t + (\rho u^i)_{,i}) Q 
+ (S_t + u^iS_{,i}) R 
\label{char}
\end{equation}
in terms of the {\it multipliers}
\begin{equation}
Q^u_i = E_{u^i}(T) , \quad 
Q^\rho= E_\rho(T), \quad 
Q^S= E_S(T) . 
\label{mult}
\end{equation}
In this formulation $u^i,\rho,S$ are given by 
{\it arbitrary} functions of $t$ and $x^i$,
with $\tilde{X}^i$ differing from $X^i$ by terms that are linear homogeneous 
in the fluid equations (\ref{eqn_1})--(\ref{eqn_3}) 
and their total spatial derivatives. 
Since the spatial Euler operators $E_{u^i}$, $E_\rho$, $E_S$ 
annihilate any spatial divergence $D_i \Theta^i$, 
there is then a one-to-one correspondence between 
nontrivial conserved densities $T$ (modulo spatial divergences) 
and non-zero multipliers $\{Q^u_i,Q^\rho,Q^S\}$. 

Necessary and sufficient equations for determining multipliers (\ref{mult})
are given by applying variational derivative operators 
$\delta/\delta u^i$, $\delta/\delta \rho$, $\delta/\delta S$
to the characteristic equation (\ref{char}),
yielding a linear homogeneous polynomial system in 
$u^i_t$, $\rho_t$, $S_t$, $u^i_{t,j}$, $\rho_{t,j}$, $S_{t,j}$, etc. 
whose coefficients must separately vanish. 
The resulting determining equations for $Q^u_i$, $Q^\rho$, $Q^S$ consist of 
the adjoint of the determining equations for symmetries (\ref{symmdeteqns}), 
plus additional integrablity equations (Helmholtz conditions) 
for $Q^u_i$, $Q^\rho$, $Q^S$ to have the variational form (\ref{mult}). 
Thus, multipliers can be characterized as 
adjoint-symmetries that have a variational form 
\cite{AB1997,AB2002a,AB2002b,BCA}.
Moreover, 
the corresponding conserved density $T$ (up to an arbitrary spatial divergence)
can be constructed explicitly from the multipliers $Q^u_i$, $Q^\rho$, $Q^S$ 
by means of homotopy integral formulas \cite{Olv,AB2002b,BCA} 
or by an algebraic scaling formula \cite{Anc}
based on invariance of the fluid equations under dilations 
$t\to \lambda t$, $x^i\to \lambda x^i$.
The determination of multipliers and hence of conservation laws 
is thereby reduced to an adjoint version of the determination of symmetries.

To conclude, in the following two tables 
we list the multipliers for, firstly,  
the non-isentropic kinematic conservation laws 
(\ref{mass})--(\ref{ener}) 
and the non-isentropic vorticity conservation laws 
(\ref{vortconslaw}) and (equivalently) (\ref{circconslaw})--(\ref{circentr}),
all of which hold for a general \eos/ (\ref{neos});
and, secondly, 
the extra non-isentropic kinematic conservation laws (\ref{similener})--(\ref{dilener})
holding only for the distinguished polytropic \eos/ (\ref{polytropic}).\\

\noindent \begin{tabular}{|c|c|c|c|c|}
\hline
Conserved density $T$ & Description & $Q^u_i=\delta T/\delta u^i$ & $Q^\rho=\delta T/\delta \rho$  & $Q^S=\delta T/\delta S$\\
\hline
$\rho$ & Mass & $1$ & $0$ & $0$\\
$\rho u^k$ & Momentum & $\rho \delta^k_i$ & $u^k$  & $0$\\
$\rho(u^jx^k - u^kx^j)$ &Angular momentum & $\rho (x^k\delta^j_i-x^j\delta^k_i)$ & $u^jx^k - u^kx^j$  & $0$\\
$\rho(tu^k-x^k)$ & Galilean momentum & $\rho t\delta^k_i$ & $tu^k-x^k$   & $0$\\
$\rho(\tf{1}{2} u^ku_k+e)$ & Energy & $\rho u_i$ & $\tf{1}{2}u^ku_k + e+\rho e_\rho$ & $\rho e_S$\\
$\rho S$ & Volumetric entropy & $0$ & $S$ & $\rho$\\
$\varpi S$ & Circulatory & $S{}_{,}{}^j \partial\varpi/\partial\omega^{ij}$ & $0$ & $\varpi$\\
%$\epsilon_{iji_1j_1\cdots i_{m-1}j_{m-1}}\omega^{i_1j_1}\cdots\omega^{i_{m-1}j_{m-1}} u^i S{}_{,}{}^j$ 
$\equiv *(\omega^{m-1})_{ij} u^i S{}_{,}{}^j$ 
& entropy $(n=2m)$ &&&\\
\hline  
\end{tabular}\\\\

\noindent \begin{tabular}{|c|c|c|c|c|}
\hline
Conserved density $T$ & Description &  $Q^u_i=\delta T/\delta u^i$ & $Q^\rho=\delta T/\delta \rho$ & $Q^S=\delta T/\delta S$\\
\hline
$tE-\tf{1}{2}\rho u^kx_k$ & Similarity Energy & 
$\rho(tu_i-\tf{1}{2} x_i)$ & $\tf{1}{2}u^k(tu_k - x_k)\qquad$ & $t e_S$\\
& & & $+ t(1+ \tf{1}{2})n\kappa(S)\rho^{2/n}$  &  \\
$t^2E-\rho (tu^k-\tf{1}{2} x^k)x_k$ & Galilean Energy & 
$t\rho(t u_i-x_i)$ & $\tf{1}{2}(tu^k - x^k)(tu_k - x_k)$ & $t^2 e_S$ \\
& & & $+t^2(1+ \tf{1}{2}n)\kappa(S)\rho^{2/n}$ &  \\	
\hline
\end{tabular}\\
Here 
$E=\rho(\tf{1}{2} u^k u_k+e) 
= \tf{1}{2}\rho u^ku_k+\tf{1}{2}n\kappa(S)\rho^{1+2/n}$ 
is the polytropic energy density,
in terms of the internal energy density $e= \tf{1}{2}n\kappa(S)\rho^{2/n}$.

\appendix
\section{Proof of Theorem~\ref{kin_class} and Proposition~\ref{kin_com}}
\label{kin_proof}

The time derivative of a kinematic conserved density $T(t,x^i,\rho,u^i,S)$ is given by 
\begin{equation}
{\mathcal D}_t T= -T_S u^i S_{,i}-T_\rho(u^i\rho_{,i}+\rho u^i{}_{,i})-T_{u^i}(u^ju^i{}_{,j}+\rho^{-1}P_,{}^i)+T_t.
\label{ntderiv}
\end{equation}  
By applying each Euler operator $E_\rho$, $E_S$, $E_{u^i}$ 
to (\ref {ntderiv}), we get three linear inhomogeneous expressions in 
$u^j{}_{,j}$, $u^j{}_{,i}$, $\rho_,{}^j$, $S_,{}^j$, 
whose coefficients must separately vanish. 
This yields the system of determining equations 
\begin{gather}
\rho^{-1}P_ST_{u^i x_i}+T_{Sx^i}u^i+T_{tS}=0, 
\label{neqn7}\\
\rho T_{\rho x^i}+T_{u^ix^j}u^j+T_{tu^i}=0, 
\label{neqn6}\\
T_{\rho x^i}u^i + \rho^{-1} P_\rho T_{u^i x_i} + T_{t\rho}=0, 
\label{neqn3}\\
(\rho T_{\rho S}-T_S)\delta_{ij}-\rho^{-1}P_S T_{u^iu^j}=0, 
\label{neqn5}\\
\rho^2 T_{\rho\rho}\delta_{ij} - P_\rho T_{u^i u^j}=0, 
\label{neqn1}\\
(\rho T_{u^i\rho}- T_{u^i})P_S - \rho P_\rho T_{u^i S} = 0, 
\label{neqn2}\\
\rho T_{u^i\rho}- T_{u^i}=0, 
\label{neqn4}
\end{gather}
to be solved for $T$.
Here $\delta_{ij} = \delta_{(ij)} \leftrightarrow \mathbf{g}$ denotes 
the Cartesian components of the Euclidean metric tensor on $\mathbb{R}^n$.

We start by integrating equation (\ref {neqn4}) with respect to $\rho$ and $u^i$, 
yielding 
\begin{equation}
T=\rho f(t,x^i,u^i,S)+ g(t,x^i,\rho , S). \label 
{neqn8}
\end{equation}
Now substituting (\ref {neqn8}) into (\ref {neqn2}) we get $f_{u^i S} = 0$ 
and then integrating gives
\begin{equation}
f = \tilde{f}(t,x^i,u^i) 
\label{neqn9}
\end{equation}
where an integration constant $c(t,x^i,S)$ has been dropped 
since we can absorb $\rho c(t,x^i,S)$ into the term $g(t,x^i,\rho,S)$ in (\ref{neqn8}).

Next, 
substituting (\ref {neqn8}) and (\ref {neqn9}) into 
both (\ref {neqn1}) and (\ref {neqn5}), 
we obtain 
\begin{gather}
\rho g_{\rho\rho}\delta_{ij}- \tilde{f}_{u^i u^j} P_\rho = 0, 
\label {neqn10} \\
(\rho g_{\rho S}-g_S)\delta_{ij}- \tilde{f}_{u^iu^j}P_S = 0. 
\label {neqn11}
\end{gather}
Separation of (\ref{neqn10}) with respect to $u^i$ yields the equations
\begin{gather}
\tilde{f}_{u^iu^j}=c(t,x^k)\delta_{ij}, 
\label{neqn13}\\
g_{\rho \rho}=c(t,x^k)\rho^{-1}P_\rho,  
\label{neqn14}
\end{gather}
where $c(t,x^k)$ is a separation constant. 
Substituting (\ref{neqn13}) into (\ref{neqn11}), we get
\begin{equation}
\rho g_{\rho S} - g_S = c(t,x^k)P_S. 
\label{neqn14a} 
\end{equation}
By writing (\ref{neqn11}) and (\ref{neqn14a}) as 
$(\rho g_{\rho} - g - cP)_\rho = (\rho g_{\rho} - g - cP)_S = 0$, 
then integrating with respect to $\rho$ and $S$, 
we obtain a first-order linear equation with respect to $\rho$, 
\begin{equation}
\rho g_{\rho} - g = c P - c_1(t,x^i). 
\label{neqn14aa}
\end{equation}
Integration of (\ref{neqn14aa}) and (\ref{neqn13}) then yields 
\begin{equation}
\tilde{f}=\tilde{c}_j(t,x^i)u^j+\tf{1}{2}c(t,x^i)u^ju_j, \quad
g= \rho^{-1}c_1(t,x^i) + c(t,x^i) \int \rho^{-2}P(\rho ,S) d\rho ,
\end{equation} 
after integration constants have been absorbed as before. 
Consequently, from (\ref{neqn8}) we have
\begin{equation}
T=c_1(t,x^i)+ c_2(t,x^i,S)\rho + \tilde{c}_j(t,x^i)\rho u^j + c(t,x^i)(\tf{1}{2}\rho u^ju_j+\rho e(\rho ,S)) 
\label{nT}
\end{equation}
which gives the general solution of the determining equations 
(\ref{neqn4}),(\ref{neqn2}),(\ref{neqn1}),(\ref{neqn5}),
with $e(\rho,S)$ given by (\ref{thermoenergy}) in terms of $P(\rho,S)$. 
Note the term $c_1$ in (\ref{nT}) is a trivial conserved density, 
and so we will put $c_1=0$.

By substituting (\ref{nT}) into the remaining determining equations 
(\ref{neqn3}), (\ref{neqn6}), (\ref{neqn7}), 
we get a system of equations to be solved for 
$c(t,x^i)$, $\tilde{c}_j(t,x^i)$, $c_2(t,x^i,S)$, $c_1(t,x^i)$:
\begin{gather}
c_{x^i}=0, 
\label{ni}\\
\tilde{c}_{jx^i}+\tilde{c}_{ix^j}+c_t\delta_{ij}=0, 
\label{nii}\\
c_{2x^i}+\tilde{c}_{it}=0, 
\label{niii}\\
\tilde{c}_{ix_i}P_\rho + c_t e + c_t \rho e_\rho + c_{2t} = 0, 
\label{niv}\\
\tilde{c}_{ix_i}P_S+c_t\rho e_S + \rho c_{2St} = 0, 
\label{nv}\\
c_{2Sx^i}=0. 
\label{nvi}
\end{gather}

To begin, 
we note (\ref{ni}) implies $c=c(t)$, and hence (\ref{nii}) has the form of 
a time-dependent homothetic Killing vector equation on $\tilde{c}_i$. 
Its solution consists of a linear polynomial in $x^i$ given by
\begin{equation}
\tilde{c}_i = C_{0i}(t)+C_{1ij}(t)x^j-\tf{1}{2}c'(t)x_i,\quad 
C_{1ij}(t)=-C_{1ji}(t). 
\label{nkv}
\end{equation}
Then, using the trace of (\ref{nii}) in (\ref{niv}), 
followed by differentiating with respect to $x^i$, we find 
\begin{equation}
c_{2tx^i} = 0. 
\label{nviia}
\end{equation}
Integration of (\ref{nvi}) and (\ref{nviia}) gives 
\begin{equation}
c_2 =  \tilde{c}_2(t,S) + \hat{c}_2(x^i). 
\label{nvii}
\end{equation}
Substituting (\ref{nvii}) and (\ref{nkv}) into (\ref{niii}), we get
\begin{equation}
\hat{c}_{2 x^i} = - C'_{0i}-C'_{1ij}x^j+\tf{1}{2}c''x_i. 
\label{nviii}
\end{equation}
By differentiating with respect to $x^j$ and antisymmetrizing in $i,j$,
we find $C'_{1ij} = 0$, 
so thus $C_{1ij}$ is constant. 
Integration of (\ref{nviii}) thus yields 
\begin{equation}
\hat{c}_2 = C_2- C'_{0i}x^i+\tf{1}{4}c''x_i x^i 
\label{nviiia}
\end{equation}
where $C_2$ is a constant. 
Then, differentiating (\ref{nviiia}) with respect to $t$, 
we get $- C''_{0i}x^i+\f{1}{4}c'''x_ix^i = 0$ 
which splits with respect to $x^i$ into $C''_{0i} = 0$, $c''' = 0$.  
Hence we obtain  
\begin{equation}
C_{0i} = a_{0i}+a_{1i}t, \quad c=b_0+b_1t+b_2t^2, 
\label{nxii}
\end{equation}
and, from (\ref{nkv}) and (\ref{nviiia}),  
\begin{equation}
\tilde{c}_i = a_{0i}+a_{1i}t + C_{1ij}x^j - \tf{1}{2}(b_1 + b_2t)x_i, \quad 
\hat{c}_2= C_2 - a_{1i}x^i +\tf{1}{2}b_2 x^i x_i,  
\label{2nxv}
\end{equation}
where $a_{0i}$,$a_{1i}$,$b_0$, $b_1$, $b_2$ are constants.
Now, through (\ref{2nxv}), we can rewrite (\ref{niv}) and (\ref{nv}) as
\begin{equation}
c'(-\tf{1}{2}nP + \rho e)_\rho = -(\rho \tilde{c}_{2t})_\rho, \quad 
c'(-\tf{1}{2}nP + \rho e)_S = -(\rho \tilde{c}_{2t})_S.
\label{niva}
\end{equation} 
Integrating (\ref{niva}) with respect to $\rho$ and $S$, we obtain
\begin{equation}
(\rho e-\tf{1}{2}nP)c'+ \rho \tilde{c}_{2t} = a(t). 
\label{nxiii}
\end{equation}
By considering the separation of (\ref{nxiii}) with respect to $\rho$, 
we have the following two cases.

\textbf{Case} $c'=0$: 
Here $c = {\rm const.} = b_0$ and hence $b_1= b_2=0$ in (\ref{nxii}).
Now splitting (\ref{nxiii}) with respect to $\rho$ yields 
$a=0$ and $\tilde{c}_{2t} = 0$, so thus 
\begin{equation}
\tilde{c}_2 = b(S) 
\label{nxvii}
\end{equation}
where, through (\ref{nvii}), 
we have absorbed an integration constant into $C_2$.
Hence, (\ref{nT}) reduces to 
\begin{equation}
T=C_2\rho + b(S)\rho + a_{0i}\rho u^i + a_{1i}\rho (t u^i - x^i)+C_{1ij}\rho u^ix^j+b_0(\tf{1}{2}\rho u^iu_i+\rho e(\rho,S))
\end{equation}
with arbitrary constants 
$a_{0i}$, $a_{1i}$, $b_0$, $C_{1ij}=-C_{1ji}$, $C_2$, 
and an arbitrary function $b(S)$, 
where $e(\rho,S)$ is given by (\ref{thermoenergy}).

\textbf{Case} $c'\neq 0$:
Dividing (\ref{nxiii}) by $\rho$, 
then differentiating with respect to $\rho$ and multiplying by $\rho^2$, 
we get 
$\rho^{2}c'(e_\rho-\f{1}{2}n(\rho^{-1}P)_\rho) = -a$
which separates with respect to $t$ and $S$ into the equations
\begin{gather}
\rho^{2}(e_\rho-\tf{1}{2}n(\rho^{-1}P)_\rho) = d, 
\label{nxix}\\
c'd = -a, 
\label{nxx}
\end{gather}
where $d$ is a constant of separation. 
Substituting $e$ from (\ref{thermoenergy}) into (\ref{nxix}), 
we obtain a first-order linear equation with respect to $\rho$, 
$-\f{1}{2}n\rho P_\rho +(1+\f{1}{2}n)P=d$, 
whose general solution is given by 
\begin{equation}
P(\rho ,S)= \kappa(S)\rho^{1+2/n} + d/(1+n/2). 
\label{sol_ODE}
\end{equation}
Since we can (without loss of generality) 
shift the pressure $P$ by an arbitrary constant, we put $d=0$. 
Then, (\ref{nxx}) yields $a=0$, 
while (\ref{thermoenergy}) becomes 
\begin{equation}
e(\rho,S)= \tf{1}{2}n\rho^{-1}P + \tilde{e}(S). 
\label{eS}
\end{equation}
Substituting (\ref{eS}) into (\ref{nxiii}), we get 
$c'\tilde{e}(S)+\tilde{c}_{2t} = 0$ 
which gives 
\begin{equation}  
c\tilde{e}+\tilde{c}_2 = b(S).
\end{equation}

As a result, (\ref{nT}) becomes 
\begin{align}
T=&
C_2\rho + b(S)\rho + a_{0i}\rho u^i + a_{1i}\rho (t u^i - x^i)
+C_{1ij}\rho u^ix^j+b_0\big(\tf{1}{2}\rho u^iu_i+\tf{1}{2}n\kappa(S)\rho^{1+2/n}\big)
\nonumber\\& 
+b_1\big(t(\tf{1}{2}\rho u^iu_i+\tf{1}{2}n\kappa(S)\rho^{1+2/n})
-\tf{1}{2}\rho x^i u_i\big)+b_2\big(t^2 (\tf{1}{2}\rho u^iu_i
+\tf{1}{2}n\kappa(S)\rho^{1+2/n})-t\rho x^i u_i\big)
\end{align}
with arbitrary constants 
$a_{0i}$, $a_{1i}$,$b_0$, $b_1$, $b_2$, $C_{1ij}=-C_{1ji}$, $C_2$, 
and an arbitrary function $b(S)$.
Also, (\ref{sol_ODE}) and (\ref{eS}) respectively reduce to 
\begin{equation}  
P(\rho ,S)= \kappa(S)\rho^{1+2/n}, \quad 
e(\rho ,S)= \tf{1}{2}n\kappa(S)\rho^{2/n}.
\end{equation}

\section{Proof of Theorem~\ref{vort_class}
and Propositions~\ref{vort_com} and~\ref{mbcom_class}}
\label{vort_proof}

To begin, we write out the component form of 
the fluid curl, the vorticity scalar and vector, 
along with their transport equations:
\begin{align}
&{\boldsymbol\omega} \leftrightarrow \omega^{ij} = 
\tf{1}{2}(u^i{}_,{}^j-u^j{}_,{}^i) = u^{[i}{}_,{}^{j]}, 
\\
&\varpi \leftrightarrow \varpi =
\epsilon^{i_1j_1\cdots i_mj_m}\omega_{i_1j_1}\cdots\omega_{i_mj_m},
\quad m=n/2,
\\
& {\boldsymbol\varpi} \leftrightarrow \varpi^i = 
\epsilon^{i j_1k_1\cdots j_mk_m}\omega_{j_1k_1}\cdots\omega_{j_mk_m}, 
\quad m=(n-1)/2, 
\\ 
&\omega^{ij}_t = (u_k\omega^{ki})_,{}^j-(u_k\omega^{kj})_,{}^i 
= u_{k,}{}^i \omega^{jk} - u_{k,}{}^j \omega^{ik} - u^k \omega^{ij}{}_{,k},
\\
& \varpi_t = -( \varpi u^i +\rho^{-1}w^{ij}P_{,j} ){}_{,i} 
= -\varpi u^i{}_{,i}- u^i \varpi_{,i}+\rho^{-2}P_S w^{jk}\rho_{,j}S_{,k}, 
\\
&\varpi^i_t = (\varpi^j u^i-\varpi^i u^j-\rho^{-1}W^{ijk}P_{,k})_{,j}
= \varpi^j u^i{}_{,j}- \varpi^i u^j{}_{,j}-u^j\varpi^i{}_{,j}+\rho^{-2}P_S W^{ijk}\rho_{,j}S_{,k}.
\end{align}
In addition, we will need the component form for
\begin{align}
&\mathbf{w} \leftrightarrow w^{ij} =
\epsilon^{iji_1j_1\cdots i_{m-1}j_{m-1}}\omega_{i_1j_1}\cdots\omega_{i_{m-1}j_{m-1}},
\quad m=n/2,\\
&\mathbf{W} \leftrightarrow W^{ijk} = 
\epsilon^{ijk j_1k_1\cdots j_{m-1}k_{m-1}}\omega_{j_1k_1}\cdots\omega_{j_{m-1}k_{m-1}}, 
\quad m=(n-1)/2, \\
&\nabla\varpi \leftrightarrow \varpi_{,i} = w^{jk} \omega_{jk}{}_{,i},\quad
\nabla{\boldsymbol\varpi} \leftrightarrow \varpi^j{}_{,i} = W^{jkl} \omega_{kl}{}_{,i}, 
\end{align}
along with the identities 
\begin{align}
& w^{(ij)} = 0, \quad 
w^{ij}{}_{,j} = 0, \quad 
n w^{ij}\omega_{kj} = \varpi\delta_k^i 
\label{id3} \\
& W^{i(jk)} = W^{(ij)k}=0, \quad
W^{ijk}{}_{,k} = 0, 
\label{id2} \\
& \varpi^i{}_{,i}=0.
\label{id1}
%\\& \varpi^i \omega_{ij}=0 . \label{id0} 
\end{align} 
Here 
$\epsilon^{i_1\cdots i_n} = 
\epsilon^{[i_1\cdots i_n]}\leftrightarrow {\boldsymbol\epsilon}$, 
$\delta_{ij} = \delta_{(ij)} \leftrightarrow \mathbf{g}$
are the Cartesian components of the spatial orientation tensor 
and the Euclidean metric tensor on $\mathbb{R}^n$; 
round brackets denote symmetrization of the enclosed indices, 
and square brackets denote antisymmetrization.

We proceed by explicitly solving the determining equations (\ref{ndeteqn})
for conserved densities of vorticity type in even and odd dimensions $n>1$.

\textbf{Case} $n=2m$: 
For a conserved density of the form $T(S,\rho,u^i,\varpi)$, 
the time derivative is given by 
\begin{equation}
{\mathcal D}_t T= 
D_i(-u^i T)
+u^i{}_{,i}A 
-\rho^{-1}(P_\rho\rho_,{}^i+P_SS_,{}^i)T_{u^i}
+\rho^{-2}P_Sw^{ij}\rho_{,i}S_{,j}T_\varpi
\label{eqn_b1}
\end{equation}
after use of identities (\ref{id3}), 
where
\begin{equation}
A= T- \rho T_\rho - \varpi T_\varpi.
\label{eqn_c1}
\end{equation}
We begin by substituting (\ref{eqn_b1}) into the determining equations 
\begin{align}
0=E_S({\mathcal D}_t T) =& 
u^{ij}\rho^{-2}P_S (\rho T_{u^i u^j}+m\rho_,{}^k w_{ik} T_{u^j\varpi} )
+ \rho_,{}^i\rho^{-2}P_S( \rho T_{u^i\rho} +\tf{1}{2}\varpi T_{u^i \varpi}
-T_{u^i}-f T_{u^iS} ) 
\nonumber \\&
+ \varpi_,{}^i\rho^{-2}P_S(\rho T_{u^i\varpi}-m w_{ij}\rho_,{}^j T_{\varpi\varpi})
+ \tr{u} A_S 
\label{eqn_ed0}
\end{align}
and 
\begin{align}
0=E_\rho({\mathcal D}_t T) =& 
u^{ij}\rho^{-2}P_S ( f T_{u^iu^j}-m S_,{}^k w_{ik} T_{u^j\varpi} )
+S_,{}^i\rho^{-2}P_S( T_{u^i}- \tf{1}{2}\varpi T_{u^i \varpi} -\rho T_{u^i\rho}) 
\nonumber \\&
+\varpi_,{}^i\rho^{-2}P_S(fT_{u^i\varpi}-m w_{ij}S_,{}^jT_{\varpi\varpi})
+ \tr{u} A_\rho 
\label{eqn_ed1}
\end{align}
where we have introduced the notation
\begin{equation}
\tr{u}=\delta_{ij} u^{ij}=u^i{}_{,i},\quad u^{ij}=\tf{1}{2}(u^i{}_,{}^j+u^j{}_,{}^i)=  u^{(i}{}_,{}^{j)},\quad f = \rho P_\rho/P_S .
\label{eqn_not}
\end{equation}
Since $T$ does not contain any derivatives of $\varpi$, $S$ and $\rho$, 
the coefficients of the separate terms 
$\varpi_,{}^iS_,{}^j$,  $\varpi_,{}^i\rho_,{}^j$, and $\varpi_,{}^i$ 
in both (\ref{eqn_ed0}) and (\ref{eqn_ed1}) must vanish, 
yielding
\begin{equation}
T_{\varpi \varpi} = 0, \quad 
T_{u^i \varpi} = 0. 
\label{eqn_ee0}
\end{equation}
Integrating (\ref{eqn_ee0}), 
and dropping a kinematic term $c(S,\rho,u^i)$ that does not involve $\varpi$, 
we get
\begin{equation}
T = a(S,\rho)\varpi.
\label{eqn_f1}
\end{equation}
Then (\ref{eqn_ed0}) and (\ref{eqn_ed1}) reduce to 
$0 = \tr{u} A_S = \tr{u} A_\rho$, 
and hence we have $0 = A_S = A_\rho$ which gives 
\begin{equation}
(\rho a_\rho)_S = (\rho a_\rho)_\rho = 0.
\end{equation}
Integration of these first-order linear equations for $a_\rho$ yields
\begin{equation}
a = b(S) + c \ln \rho ,\quad
c={\rm const.} . 
\label{eqn_h1}
\end{equation}

Thus from (\ref{eqn_f1}) and (\ref{eqn_h1}) we have
\begin{equation}
T = b(S)\varpi + c\varpi \ln \rho
\label{eqn_i10}
\end{equation}
and
\begin{equation}
{\mathcal D}_t T = 
D_i(-u^i T) -c\varpi u^i{}_{,i} + (b(S)+c\ln\rho)\rho^{-2}P_S w^{ij}\rho_{,i}S_{,j}.
\label{eqn_j1}
\end{equation}
The third term in (\ref{eqn_j1}) can be written as a spatial divergence 
$-D_i(B(S,\rho)w^{ij}\rho_{,j}) = w^{ij}B_S\rho_{,i}S_{,j}$ 
through the identities (\ref{id3}), 
where $B(S,\rho)$ satisfies $B_S = \rho^{-2}(b(S)+c\ln\rho)P_S$. 
Thus, (\ref{eqn_j1}) becomes 
\begin{equation}
{\mathcal D}_t T = -D_i(u^i T + Bw^{ij}\rho_{,j})- c\varpi u^i{}_{,i}.
\label{eqn_j0}
\end{equation}
Now we substitute (\ref{eqn_j0}) into the final determining equation
\begin{equation}
0=E_{u^i}({\mathcal D}_t T) = c\varpi_{,i}+mc\tr{u}_,{}^j w_{ji} 
\label{eqn_l10}
\end{equation}
Since $\varpi$ has no dependence on $u^{ij}$ 
(which is linearly independent of $\omega^{ij}$), 
we obtain
\begin{equation}
c=0.
\label{eqn_m1}
\end{equation}

Therefore, (\ref{eqn_m1}) and (\ref{eqn_i10}) give the result
\begin{equation}
T = b(S)\varpi
\label{eqn_n1}
\end{equation}
with
\begin{equation}
{\mathcal D}_t T = -D_i(u^i b(S)\varpi + B(S,\rho)w^{ij}\rho_{,j}) ,\quad
B(S,\rho) = \rho^{-2}\int b(S)P_S(S,\rho)dS .
\label{eqn_o1}
\end{equation}
We now note that if $b={\rm const.}$ then (\ref{eqn_n1}) is 
a trivial conserved density
\begin{equation}
T= b\varpi =b(w^{ij} u_j)_{,i} = D_i(b w^{ij} u_j).
\label{eqn_p1}
\end{equation}
Its corresponding moving-flux $\xi^i = b\rho^{-2}P w^{ij}\rho_{,j}$
is non-vanishing due to the form of (\ref{eqn_o1}).

This completes the proof in the even-dimensional case.

\textbf{Case} $n=2m+1$: 
For a conserved density $T(S,\rho,u^i,\varpi^i)$, 
similarly to the previous case, its time derivative is given by 
\begin{equation}
{\mathcal D}_t T= 
D_i(-u^i T)
+u^i{}_{,i}(T-\rho T_\rho - \varpi^j T_{\varpi^j})
- \rho^{-1}(P_\rho\rho_,{}^i +P_SS_,{}^i)T_{\rho u^i}
+\rho^{-2}P_SW^{ijk}\rho_{,j}S_{,k}T_{\varpi^i}.
\label{eqn_b}
\end{equation}
The determining equations 
$0=E_S({\mathcal D}_t T)$ and  $0=E_\rho({\mathcal D}_t T)$ 
then yield 
\begin{align}
0 =& 
\varpi^i{}_,{}^j\rho^{-2}P_S(\rho T_{u^j\varpi^i}+\rho_,{}^l W_{il}{}^k T_{\varpi^k \varpi^j})
+ \rho^{-2}P_S\rho_,{}^i(\rho T_{u^i\rho}-T_{u^i}-fT_{u^iS}-W_{ij}{}^k\omega^{jl} T_{u^l\varpi^k})
\nonumber\\&
+ u^{ij}( \varpi_j T_{\varpi^i S} + \rho^{-1}P_S \rho T_{u^i u^j}+\rho^{-2}P_S \rho_,{}^l W_{il}{}^k T_{\varpi^k u^j})
+\tr{u}(T_S-\rho T_{\rho S}- \varpi^i T_{\varpi^i S}) 
\label{eqn_d0}
\end{align}
and 
\begin{align}
0 =& 
\varpi^i{}_,{}^j\rho^{-2}P_S(fT_{u^j\varpi^i}-S_,{}^l W_{il}{}^k T_{\varpi^k \varpi^j})
+ \rho^{-2}P_S S_,{}^i(fT_{u^iS}+T_{u^i}-\rho T_{u^i\rho}+W_{ij}{}^k\omega^{jl} T_{u^l\varpi^k})
\nonumber\\&
+ u^{ij}( \varpi_j T_{\varpi^i \rho} + \rho^{-1}P_\rho T_{u^i u^j}-\rho^{-2}P_S S_,{}^l W_{il}{}^k T_{\varpi^k u^j})
-\tr{u}(\rho T_{\rho \rho}+ \varpi^i T_{\rho \varpi^i}) 
\label{eqn_d1}
\end{align} 
using the notation (\ref{eqn_not}). 
Since $T$ does not contain any derivatives of $\varpi^i$, 
the first term in both (\ref{eqn_d0}) and (\ref{eqn_d1}) must vanish 
modulo the identity (\ref{id1}). 
This implies 
\begin{align}
& T_{\varpi^j \varpi^k} = 0, 
\label{eqn_e0} \\
& T_{u^i \varpi^j} = a(\rho,u^k,\varpi^k)\delta_{ij}. 
\label{eqn_e}
\end{align}
Applying the derivative operator $\p_{u^k}$ to (\ref{eqn_e}), 
antisymmetrizing in $[jk]$, and taking the trace over $(ij)$, 
we get 
$(n-1)a_{u^k} = 0$. 
Hence, in $n>1$ dimensions, 
\begin{equation}
a_{u^k} = 0.
\label{eqn_f}
\end{equation}
By also applying the derivative operator $\p_{\varpi^k}$ to (\ref{eqn_e}), 
we similarly get
\begin{equation}
a_{\varpi^k} = 0.
\label{eqn_g}
\end{equation}
Integration of (\ref{eqn_e0}), (\ref{eqn_e}), (\ref{eqn_f}), (\ref{eqn_g}) 
then yields
\begin{equation}
T = a(S,\rho)u^i\varpi_i + b_i(S,\rho )\varpi^i.
\label{eqn_h}
\end{equation}
Here we have dropped an integration constant $c(S,\rho,u^i)$ 
since it does not involve $\varpi^i$ (i.e. it is of kinematic form).

The determining equations (\ref{eqn_d0}) and (\ref{eqn_d1}) 
thereby reduce to
\begin{align}
& 0 = u^{ij}B_{iS}\varpi_j-\rho \tr{u}\varpi^iB_{iS\rho} + \rho_,{}^i\varpi_i \rho^{-2}A, 
\label{eqn_i1} \\
& 0 = u^{ij}B_{i\rho}\varpi_j - \tr{u} \varpi^i (\rho B_{i\rho\rho} + B_{i\rho}) - S_,{}^i\varpi_i \rho^{-2}A, 
\label{eqn_i2}
\end{align}  
with coefficients
\begin{align}
& A = (\rho a_{\rho} - fa_S - (m+1)a)P_S, 
\label{j1}\\
& B_i = a u_i + b_i. 
\label{j2}
\end{align}
In (\ref{eqn_i1}) and (\ref{eqn_i2}) the respective coefficients of 
the terms $\rho_,{}^i$ and $S_,{}^i$ must vanish, 
which yields $A=0$.
Since $P_S \not\equiv 0$, we get 
\begin{equation}
0 = \rho a_\rho -f a_S - (m+1)a.
\label{eqn_6}
\end{equation}
Then, since the coefficient of $u^{ij}$ must vanish in both (\ref{eqn_i1}) and (\ref{eqn_i2}), we obtain 
\begin{align}
& \varpi_{(j}B_{i)S} =\varpi^k \rho B_{k\rho S} \delta_{ij}, 
\label{eqn_k} \\
& \varpi_{(j}B_{i)\rho} =\varpi^k (\rho B_{k\rho})_\rho \delta_{ij} . 
\label{eqn_k1}
\end{align}
By taking the product of (\ref{eqn_k}) with $\varpi_l \varpi_h$ 
and antisymmetrizing in $[ih]$ and $[jl]$, 
we find
$\rho\varpi^k B_{k\rho S}\varpi_{[h}\delta_{i][j} \varpi_{l]} = 0$. 
This implies $B_{k\rho S} =0$. 
The same antisymmetric product applied to (\ref{eqn_k1}) 
then implies $(\rho B_{k\rho})_\rho =0$. 
Hence (\ref{eqn_k}) and (\ref{eqn_6}) become 
$\varpi_{(j} B_{i)S} = \varpi_{(j}B_{i)\rho} = 0$, 
which yields
\begin{equation}
B_{iS} = B_{i\rho} = 0.
\label{a0}
\end{equation}
From (\ref{j2}) and (\ref{eqn_6}) we thus obtain 
\begin{equation}
a = 0 ,\quad
b^i= {\rm const.} . 
\label{eqn_l}
\end{equation}

Hence (\ref{eqn_h}) reduces to 
\begin{equation}
T = b_i \varpi^i = D_j(b_i W^{ijk} u_k) 
\label{eqn_m}
\end{equation}
which is a trivial conserved density, 
with 
\begin{equation}
{\mathcal D}_t T = D_i (b_j (2u^{[j}\varpi^{i]}-W^{ijk} (\rho^{-1})_{,k}))
\label{eqn_n}
\end{equation}
as given by (\ref{eqn_b}) combined with the identities  
$u^i{}_{,j} \varpi^j = D_j (u^i \varpi^j)$ 
and 
$\rho^{-2}P_S W^{ijk} \rho_{,j} S_{,k} = -W^{ijk} (\rho^{-1})_{,j} P_{,k} = D_k(-W^{ijk} P(\rho^{-1})_{,j})$
obtained via (\ref{id1}) and (\ref{id2}). 
The corresponding moving-flux has the form 
$\xi^i = -b_j(u^j\varpi^i +\rho^{-2} W^{ijk} \rho_{,k})$
which is non-vanishing. 

This completes the proof in the odd-dimensional case.


\begin{thebibliography}{99} 

\bibitem{Anc}
Anco, S.C.,
{Conservation laws of scaling-invariant field equations, J. Phys. A: Math. Gen., 36 (2003), 8623--8638.}

\bibitem{AB1997}
Anco, S.C. and Bluman, G.,
{Direct construction of conservation laws from field equations, Phys. Rev. Lett. 78 (1997), 2869--2873.}

\bibitem{AB2002a}
Anco, S.C. and Bluman, G.,
{Direct construction method for conservation laws of partial differential equations. Part I: Examples of conservation law classifications, 
Eur. J. Appl. Math. 13 (2002), 545--66.}

\bibitem{AB2002b}
Anco, S.C. and Bluman, G.,
{Direct construction method for conservation laws of partial differential equations. Part II: General treatment, 
Eur. J. Appl. Math. 13 (2002), 567--85.}

\bibitem{paperI}
Anco, S.C. and Dar, A., 
Classification of conservation laws of compressible isentropic fluid flow in $n>1$ spatial dimensions,
Proc. Roy. Soc. A 464 (2009), 2461--2488.

\bibitem{paperIII}
Anco, S.C. and Dar, A., 
Lower-degree conservation laws in fluid flow. 
In preparation. 

\bibitem{Arn1966}
Arnold, V.I.,
{Sur la g\'eom\'etrie diff\'erentielle des groupes de Lie de dimension infinie et ses applications \`a l'hydrodynamique des fluides parfaits, Ann. Inst. Fourier 16 (1966), 316--361.}

\bibitem{Arn1969}
Arnold, V.I.,
{The Hamiltonian nature of the Euler equation in the dynamics of rigid body and of an ideal fluid, Uspekhi Mat. Nauk 24 (1969), No. 3, 225--226.}

\bibitem{ArnKhe}
Arnold, V.I. and Khesin, B.A.,
{\it Topological Methods in Hydrodynamics}, Springer-Verlag, 1998.

%\bibitem{Batch}
%Batchelor, G.K., {\it An Introduction to Fluid Dynamics}, Cambridge University Press, 2000.

%\bibitem{BA}
%Bluman, G., Anco, S.C., {\it  Symmetry and Integration Methods for Differential Equations}, Springer, 2002.

\bibitem{BCA}
Bluman, G., Cheviakov, A., Anco, S.C.,
{\it  Applications of Symmetry Methods to Partial Differential Equations}, Springer, 2009.

\bibitem{BTA}
Bluman, G., Temerchaolu, Anco, S.C.,
New conservations laws obtained directly from symmetry action on a known conservation law,
J. Math. Anal. Appl. 322 (2006), 233--250.

%\bibitem{chorin}
%Chorin, A.J. and Marsden, J.E., {\it A Mathematical Introduction to Fluid Mechanics}, Springer-Verlag, 1997.

\bibitem{Dez}
Dezin, A.A., 
Invariant forms and some structure properties of the Euler equations of hydrodynamics, 
Zeit. Anal. Anwend. (in Russian) 2 (1983), 401--409.

%\bibitem{GS}
%Guillemin, V. and Sternberg, S., The moment map and collective motion, Ann. Phys. 127 (1980), No. 1, 220--253.

\bibitem{Ibr1973}
Ibragimov, N.H.,
Conservation laws in hydrodynamics, Dokl. Akad. Nauk USSR, 210, No. 6: 1307--1309, 1973. English transl., Soviet Physics Dokl., 18 (1973--1974).

\bibitem{Ibr}
Ibragimov, N.H.,
{\it CRC Handbook of Lie Group Analysis of Differential Equations Vol. 1,2,3}, CRC Press, 1994--1996.

\bibitem{KheChe}
Khesin, B.A. and Chekanov Y.V.,
Invariants of the Euler equations for ideal or barotropic hydrodynamics and superconductivity in D dimensions, Physica D 40 (1989), 119--131.

\bibitem{Kup}
Kupershmidt, B.A., 
{\it The Variational Principles of Dynamics},
Advanced Series in Mathematical Physics vol. 13, 
World Scientific, 1992.

\bibitem{LanLif}
Landau, L.D. and Lifshitz, E.M.,
{\it Fluid Mechanics}, Pergamon, 1968.

\bibitem{Lig}
Lighthill, M.J., 
{\it An informal introduction to theoretical fluid mechanics}, 
Oxford University Press, 1986. 

%\bibitem{MB}
%Majda, A.J. and Bertozzi A.L., {\it Vorticity and Incompressible Flow}, Cambridge University Press, 2002.

%\bibitem{MRW}
%Marsden, J., Ratiu, T., Weinstein, A., Semidirect product and reduction in mechanics, Trans. Am. Math. Soc. 281 (1984), 147--177.

%\bibitem{Nov}
%Novikov, S.P., The Hamiltonian formalism and a many-valued analogue of Morse theory, Rus. Math. Surveys 37 (1982), no. 5, 1--56.

\bibitem{Olv}
Olver, P.J.,
{\it Applications of Lie Groups to Differential Equations}, Springer-Verlag, 1993.

\bibitem{OlvNut}
Olver, P.J. and Nutku, Y.J., 
Hamiltonian structures for systems of hyperbolic conservation laws, 
J. Math. Phys. 29 (1988), 1610--1619. 

\bibitem{Ovs}
Ovsyannikov, L.V.,
{\it Group properties of differential equations} (Russian), USSR Academy of Sciences, Novosibirsk, 1962. 

\bibitem{Ser}
Serre, D., 
Invariants et d\'eg\'en\'erescenc symplectique de l'\'equation d'Euler des fluids parfaits incompressibles, 
C.R. Acad. Sci. Paris, S\'er. A 298 (1984), 349. 

\bibitem{Ver1}
Verosky, J.,
Higher-order symmetries of the compressible one-dimensional isentropic
fluid equations, 
J. Math. Phys. 25 (1984), 884--888.

\bibitem{Ver2}
Verosky, J.,
First-order conserved densities for gas dynamics, 
J. Math. Phys. 27 (1986), 3061--3063.

\bibitem{Ver}
Verosky, J.,
The Hamiltonian structure of generalized fluid equations, Lett. Math. Phys. 9 (1985), 51--53.

\bibitem{Whi}
Whitham, G.B., 
{\it Linear and nonlinear waves}, Wiley, 1974. 


\end{thebibliography}
\end{document}